\documentclass[fleqn,12pt,twoside]{article}
\usepackage{espcrc1}

\usepackage{graphicx}

\newcommand{\about}{$\simeq$}
\def\degree{\relax\ifmmode^\circ \else $^\circ$\fi}
\newcommand{\Msol}{M$_{\odot}$\ }
\newcommand{\Al}{$^{26}$Al\ }
\newcommand{\Fe}{$^{60}$Fe\ }
\newcommand{\Na}{$^{22}$Na\ }
\newcommand{\Ni}{$^{56}$Ni\ }
\newcommand{\Ti}{$^{44}$Ti\ }
\def\flux{{\hbox{ph {cm}$^{-2}$ {s}$^{-1}$}}}

\def\Be{{\douzesf  $^{\douzesf  7}$Be}}

\def\Be{$^{7}$Be\ }
\def\Al{$^{26}$Al\ }
\def\Mg{$^{26}$Mg\ }
\def\Ni{$^{56}$Ni\ }

\def\Na{$^{22}$Na\ }
\def\Co{$^{56}$Co\ }
\def\Ci{$^{57}$Co\ }

\def\Fe{$^{60}$Fe\ }

\def\Ti{$^{44}$Ti\ }

\def\ga{$\gamma$}

\def\ms{M$_{\odot}$}

\title{Astrophysical constraints from gamma-ray spectroscopy
\footnote{\it Nucl.Phys.A Special Volume on Nuclear Astrophysics,
 Eds. K.-H. Langanke, F.-K. Thielemann, M. Wiescher
 }
}
\author{Roland Diehl\address[MPE]{Max-Planck-Institut f\"ur extraterrestrische Physik,
              D-85741 Garching, Germany},
        Nikos Prantzos\address[IAP]{Institut d'Astrophysique,
              F-75014 Paris, France}, and
        Peter von Ballmoos\address[CESR]{Centre d'\'Etude Spatiale des Rayonnements,
              F-31028 Toulouse, France}
        }%
\begin{document}
\maketitle
\begin{abstract}
Gamma-ray lines from cosmic sources provide unique isotopic information, since
they originate from energy level transitions in the atomic nucleus. 
Gamma-ray telescopes explored this astronomical window
in the past three decades, detecting
radioactive isotopes that have been
ejected in interstellar space by cosmic nucleosynthesis events
and nuclei that have been excited through 
collisions with energetic particles.
Astronomical gamma-ray telescopes feature standard detectors of nuclear 
physics, but have to be surrounded by effective shields 
against local instrumental background,
and need special detector and/or mask arrangements 
to collect imaging information. Due to exceptionally-low
signal/noise ratios, progress in the field has been slow compared with
other wavelengths. Despite the difficulties, this young field
of astronomy is well established now, in particular due to advances 
made by the Compton Gamma-Ray Observatory in the 90ies.
The most important achievements so far concern:
short-lived radioactivities that have been detected in  
a couple of supernovae
($^{56}$Co and  $^{57}$Co in SN1987A,  $^{44}$Ti in Cas~A), 
the diffuse glow of long-lived  $^{26}$Al
that has been mapped along the entire plane of the Galaxy, 
several excited nuclei that have been detected 
in solar flares, and, last but not least, positron annihilation that
has been observed in the inner Galaxy since the 70ies.
High-resolution spectroscopy is now being performed: 
Since 2002, ESA's INTEGRAL and NASA's RHESSI, two 
space-based gamma-ray telescopes with Ge detectors, are in operation.
Recent results include: Imaging and line shape measurements of e$^-$-e$^+$
annihilation  emission from the Galactic bulge, which can hardly be
accounted for by conventional sources of positrons; 
$^{26}$Al emission and line width measurement from the inner Galaxy 
and from the Cygnus region, which can constrain the properties of the 
interstellar medium; and
a diffuse $^{60}$Fe gamma-ray line emission 
which appears rather weak, 
in view of current theoretical predictions.
Recent Galactic core-collapse supernovae are studied through \Ti radioactivity, 
but, apart from Cas~A, no other source  has been found; this is
a rather surpising result, assuming a canonical Galactic 
supernova rate of $\sim$1/50 years. 
The characteristic signature of \Na-line emission from a nearby 
O-Ne-Mg  novae is expected to be measured during INTEGRAL's lifetime.
\end{abstract}

\section{OVERVIEW}
  
Radioactive isotopes are common by-products of nucleosynthesis in cosmic 
sources and constitute important probes of the underlying 
physical processes, as they 
can be studied through their characteristic gamma-ray emission. 
Candidate sources are supernovae and novae, but also the winds from massive
stars and intermediate stars on the Asymptotic Giant Branch (AGB stars).
Similarly, collisions of nuclei that have been accelerated to cosmic-ray
energies produce de-excitation gamma-rays, which can be used
to study the physics of the acceleration sites, as in the case of solar flares. 

Radioactive nuclei  are thermonuclearly
synthesized in the hot and dense stellar interiors, which are opaque to
\ga-rays. Released \ga-ray photons interact with the surrounding material 
and are 
Compton-scattered down to X-ray energies, until they are photoelectrically
absorbed and their energy is emitted at longer wavelengths. To become 
detectable,
radioactive nuclei have to be brought to the stellar surface (through vigorous
convection) and/or ejected in the interstellar medium, either through stellar
winds (Asymptotic Giant Branch and Wolf-Rayet stars) or through an 
explosion (novae or supernovae). Their direct
detection provides then unique information on their production sites,
which cannot be obtained through observations at other wavelengths.
Thus it complements other methods used for the study of cosmic nucleosynthesis, 
which are often indirect or  plagued with instrumental difficulties.
Those methods exploit X-ray/low-energy gamma-ray continuum emission arising from 
the effect of Comptonization,
characteristic X-ray recombination lines from highly-ionized species, and
laboratory isotopic analysis of presolar grains included in meteorites.

\begin{table*}[t]
\centering\
\caption{Important stellar radioactivities for gamma-ray line astronomy
\label{fig:Table1}}
\includegraphics[width=0.85\textwidth]{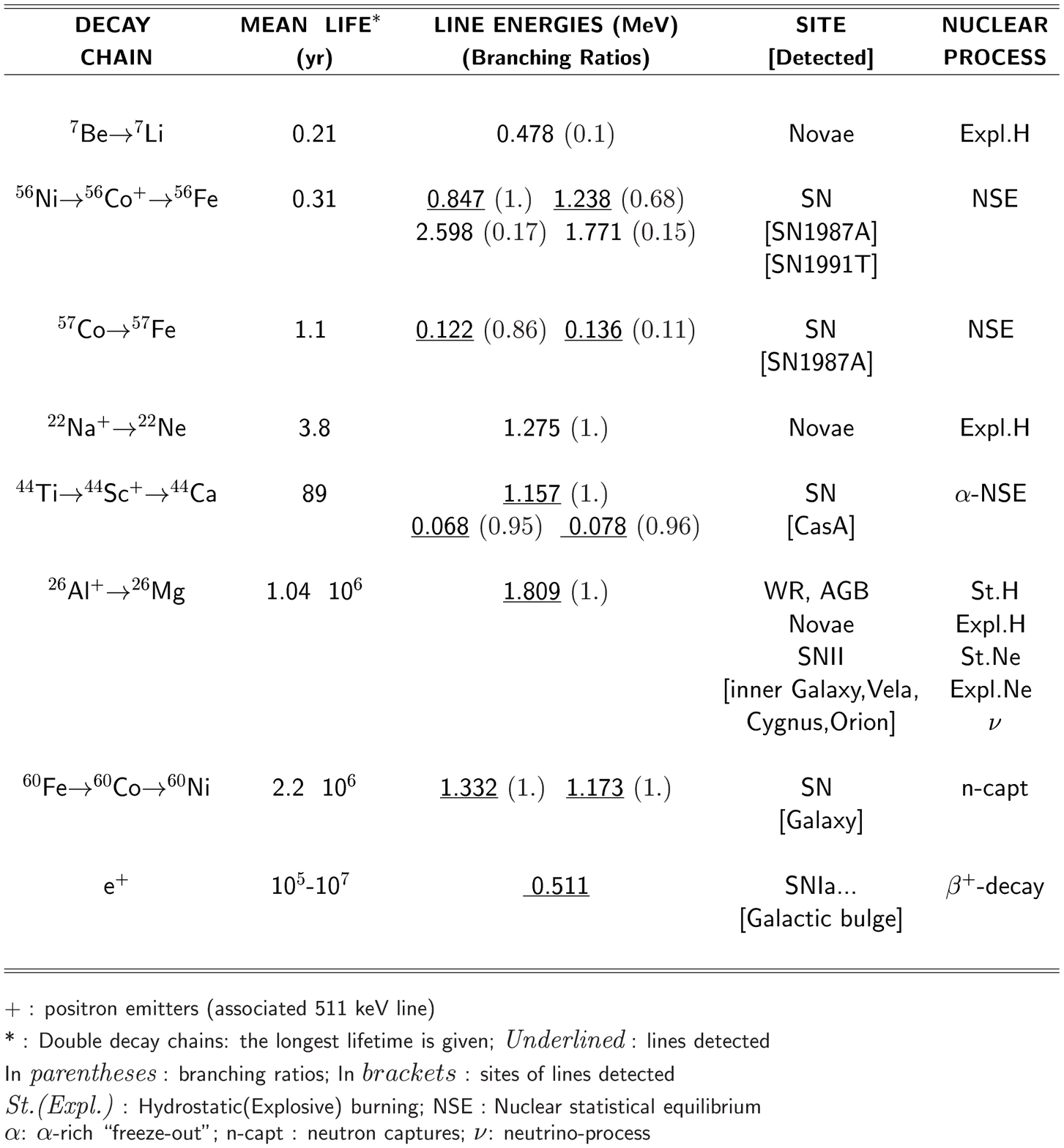}
\end{table*}

Obviously, radionuclides of interest for \ga-ray line astronomy are those
with high enough yields and short enough lifetimes for the emerging \ga-ray lines
to be detectable. On the basis of those criteria, Table 1 gives the most important
radionuclides (or radioactive chains) for \ga-ray line astronomy, along with
the corresponding lifetimes, line energies and branching ratios, production sites
and nucleosynthetic processes.


When the lifetime of a  radioactive nucleus  is not very large w.r.t.
the timescale between two nucleosynthetic events in the Galaxy,
those events are expected to be seen as {\it point-sources} in the light
of that radioactivity. In the opposite case a {\it diffuse emission} in
the Galaxy is expected from the cumulated emission of hundreds
or thousands of sources.  Characteristic timescales between two
explosions  are $\sim$1-2 weeks for novae 
(from their estimated Galactic frequency of
$\sim$25--30 yr$^{-1}$ \cite{matt03},
$\sim$50-100 yr for supernova types II+Ib, and $\sim$200-400 yr
for SNIa (from the corresponding Galactic frequencies of 
$\sim$3~SNII+SNIb century$^{-1}$ and $\sim$0.25-0.5 SNIa century$^{-1}$,
\cite{Mannucci04}).
Comparing those timescales to the decay lifetimes of Table~1,
one sees that in the case of long-lived \Al  and \Fe 
{\it diffuse} emission is expected; the spatial profile of such emission
should reflect the Galactic distribution of the
underlying sources, if the ejected nuclei do not travel too
far away from their sources during their radioactive lifetime.  
All the other radioactivities of Table 1
should be seen as {\it point sources} in the Galaxy, except, perhaps,
$^{22}$Na from Galactic novae in the central bulge. Indeed, the most prolific \Na 
producers, O-Ne-Mg rich novae, have a
frequency $\sim$1/3 of the total (i.e. $\sim$10 yr$^{-1}$),
resulting in $\sim$40 active sources in the Galaxy during
the 3.8~yr lifetime of $^{22}$Na.

In principle, the intensity of the escaping \ga-ray lines gives important
information on the yields of the corresponding isotopes, on the
physical conditions (temperature, density, neutron excess etc.) in the
stellar zones of their production, and on other features of the
production sites (extent of convection, mass loss, hydrodynamic instabilities,
position of the ``mass-cut'' in core-collapse SN, etc.).
Moreover, the shape of the \ga-ray lines reflects
the velocity distribution of the ejecta, modified by the opacity along
the line of sight and can give information on the structure
of the ejecta, and on the interstellar medium surrounding sources of
nucleosynthesis.

At this point, the main advantages of the study of nucleosynthesis through
the detection of the characteristic $\gamma$-ray lines of radioactivities should be
clear: 
\begin{itemize}
\item{}
The unique possibility to {\it unambiguously identify
isotopes}, which are the direct products of nuclear reactions. Indeed,
elementary abundances (usually revealed by observations in most
other wavelengths) may give ambiguous messages, since they may be the sum of
different isotopes, 
produced by different processes in different physical 
conditions. Isotopic measurements from presolar grains may be affected 
by various physico-chemical effects, therefore astrophysical interpretations 
depend on models for such processes.
In contrast, radioactive decay in interstellar space is mostly 
unaffected by 
physical conditions in/around the source such as temperature or density
(except for the case of electron-capture radioactivities, see Sec. 3.1 
for the case of \Ti).
\item{}
Decay gamma-rays are not attenuated along the line-of-sight 
due to their {\it highly penetrating nature}
(attenuation length \about~few g~cm$^{-2}$), thus probing stellar regions 
which are not accessible at other wavelengths.
\end{itemize}

The measurement of such decay gamma-rays with satellite-borne 
telescopes occurs in near-earth space, above the Earth's atmosphere
(which is optically thick to gamma-rays, and is by itself a bright
gamma-ray source from cosmic-ray interactions). The technique of gamma-ray 
telescopes is complex \cite{Seeon03}, and still in specific aspects
less precise than alternative isotopic abundance measurements. 
With present-day spatial resolutions of the order of a degree it cannot compete with
current X-ray telescopes such as Chandra and XMM-Newton to, e.g., map
the \Ti distribution within the Cas A supernova remnant. Furthermore,
local background from activation of the spacecraft and instrument
through their irradiation with cosmic ray particles is high, leading to
signal-to-background ratios which are of the order of 1/100; 
this is orders of magnitude 
worse than in laboratory measurements on presolar grain abundances,
and effectively limits the sensitivity to \ga-ray fluxes 
of a few 10$^{-6}$ ph~cm$^{-2}$s$^{-1}$ for realistic observing times.  
Therefore, present-day instruments can access only sources 
in our Galaxy (and up to 10 Mpc in the case of strong \Co ~lines from SNIa), 
which are sufficiently bright for $\gamma$-ray line measurements.
On the other hand, fields of view are of the order of sr, very much larger than
in X-ray and even more in optical/IR telescopes; this allows for all-sky 
mapping and monitoring
to an extent which is often impossible in these other fields of astronomy.


In the past three decades, various gamma-ray telescopes 
have established the following major features in the field
of gamma-ray line astronomy \cite{dieh98,prantzos_iws04}:

\begin{itemize}
\item{}
Interstellar \Al has been mapped along the plane of the Galaxy,
confirming that nucleosynthesis is an  ongoing process in the Milky Way 
\cite{dieh95,pran96,plue01}; quite recently, the detection of \Fe has
also been reported, bringing complementary (and poorly understood yet)
information.
\item{}
Characteristic Co decay gamma-ray lines have been observed from SN1987A
\cite{tuel90,matz88,kurf92}, directly
confirming core-collapse supernova production of fresh isotopes 
belonging to the iron group. A marginally significant signal
from Co decay  has been reported in the case of the thermonuclear 
supernova SN1991T (\cite{morr97}).
\item{}
\Ti gamma-rays have been discovered \cite{iyud94,scho00}
from the young supernova remnant Cas A,
confirming models of explosive nucleosynthesis in core-collapse supernovae.
\item{}
A diffuse glow of positron annihilation gamma-rays has been recognized
from the direction of the inner Galaxy \cite{purc97,kinz01}; its intensity 
appears to be only marginally consistent with nucleosynthetic
production of $\beta^+$-decaying radioactive isotopes. 
\end{itemize}

In the following, we discuss those issues in some detail, after a brief introduction
to the relevant theoretical background in the next section. 


\section{PRODUCTION SITES FOR GAMMA-RAY EMITTING ISOTOPES}

Most of the radioactivities in Table 1 are synthesised in supernovae,
either in core-collapse (ccSNe) or in thermonuclear (SNIa) explosions. This is the
case, in particular, for the isotopes of the Fe-peak, \Ti, \Ni and \Ci, which
are produced by explosive Si-burning in the innermost stellar layers. 
Long-lived  \Fe and \Al are produced in massive stars both hydrostatically and
explosively. In fact, \Al may also be produced in AGB stars and novae, but its
distribution in the Milky Way (as mapped by the COMPTEL instrument aboard 
CGRO) suggests that those sources are minor contributors
galaxywide (see Sec. 3.3). Finally, novae are expected to produce astrophysically 
interesting amounts of several radioactive light-element isotopes, in particular 
$^{7}$Be and $^{22}$Na.

\begin{figure}
\centering\
\includegraphics[angle=-90,width=.48\textwidth]{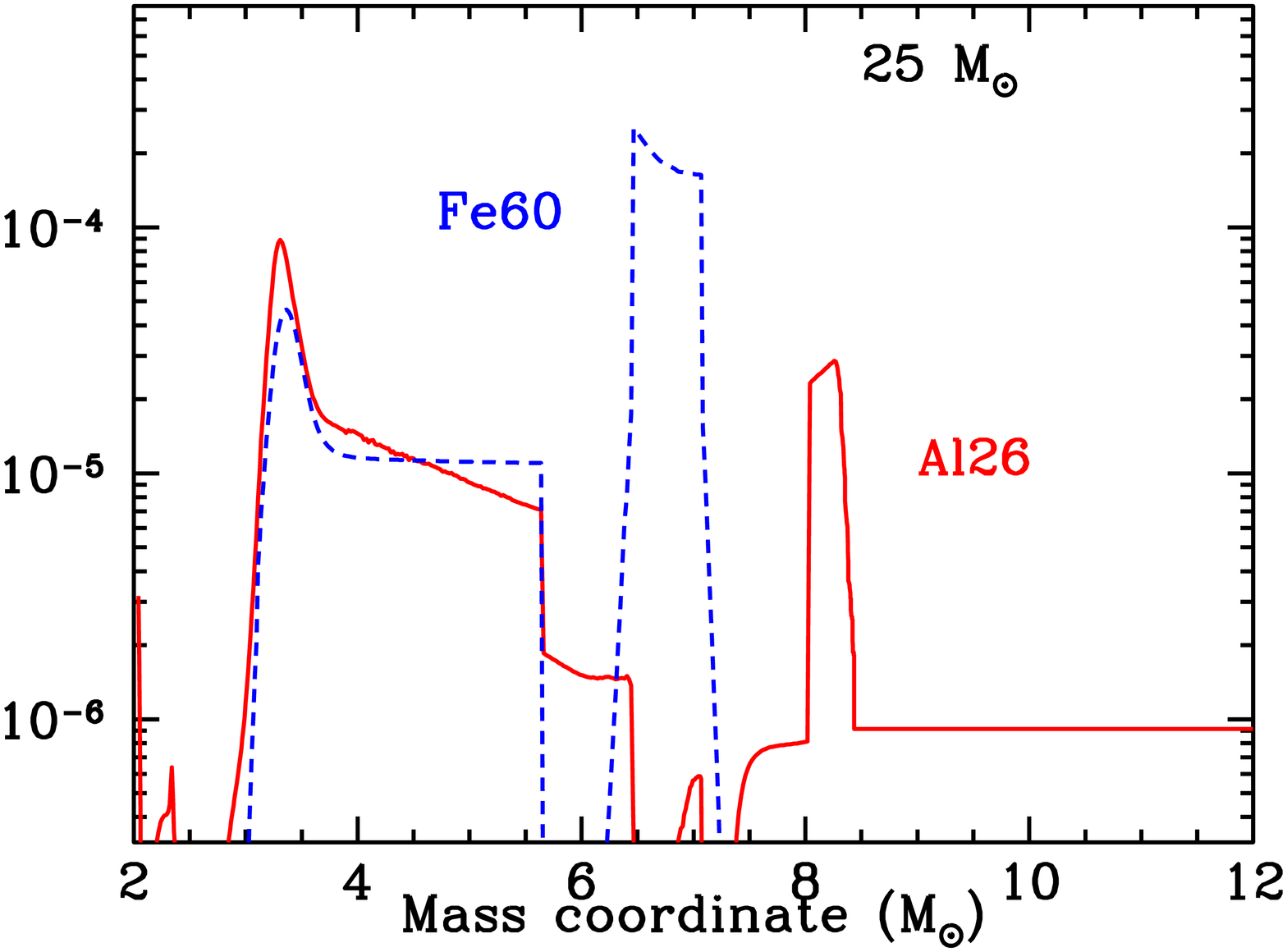}
\includegraphics[angle=-90,width=.48\textwidth]{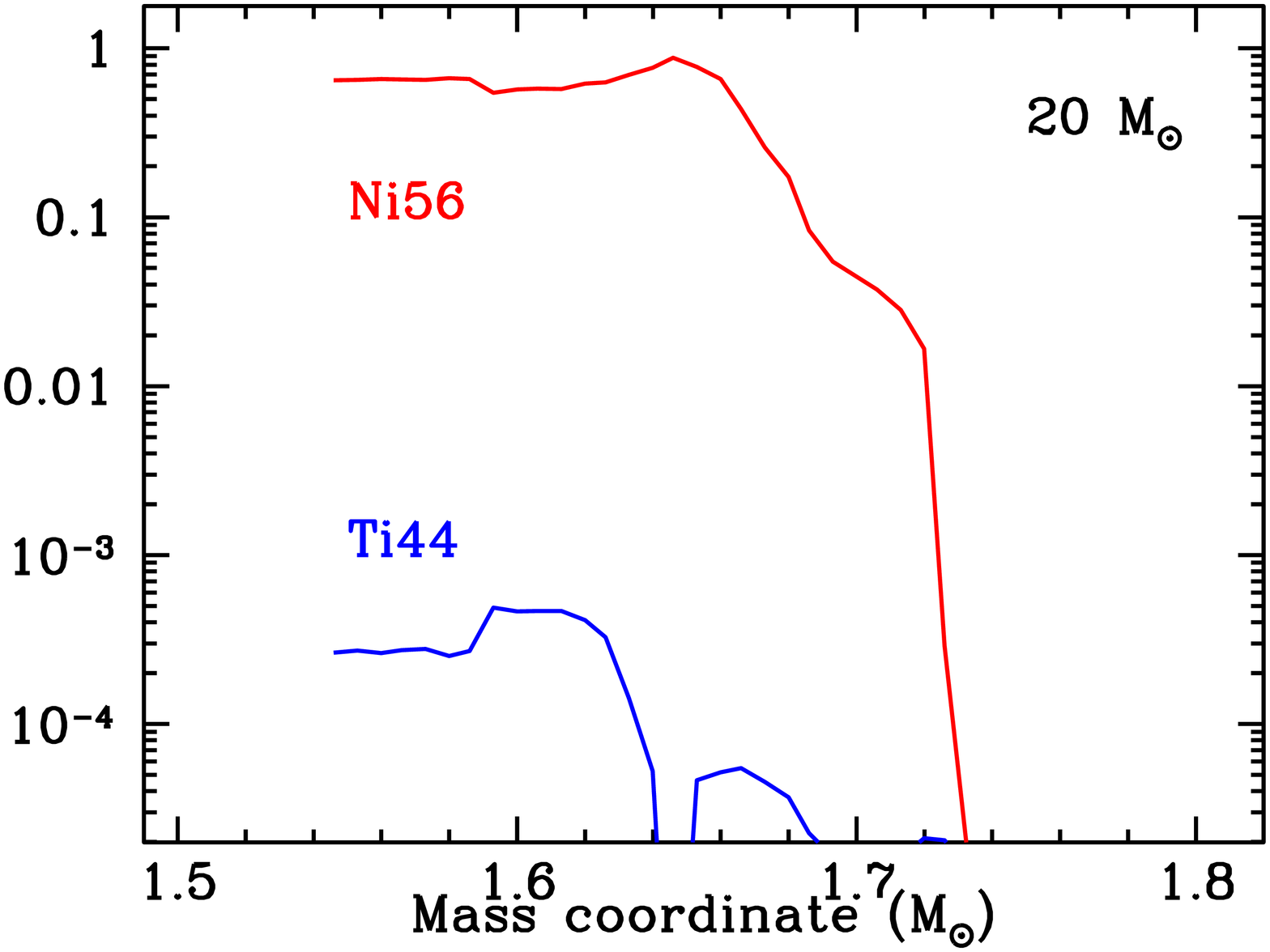}
 \caption{Radial profiles of major radioactivities in core collapse supernovae
\cite[from]{raus02}. 
 {\it Left}:
Abundance profiles (mass fractions)
of \Al (solid curve) and \Fe (dashed curve) inside
an exploded 25 \ms \ star; 
they are produced both hydrostatically (in the H- and 
He- layers, for  \Al and  \Fe, respectively)
and explosively (both nuclei in the Ne-O layers).
{\it Right}: Profiles of  \Ni and  \Ti 
inside an exploded 20 \ms \ star; the uncertain position of the ``mass-cut'' 
(around 1.5 \ms) makes
difficult an accurate prediction of their yields.
}
 \label{radioact_profiles}
\end{figure}

\Al is produced by proton captures on $^{25}$Mg, hence in stellar zones where
either of the two reactants is abundant. This may happen either
in the H-layers (H-burning core or shell) where protons are abundant, or in the
Ne-O layers, where $^{25}$Mg nuclei are abundant from Ne-burning reactions
(Fig. \ref{radioact_profiles}).
\Al produced in the H-core decays with its 1~Myr-lifetime, and is ejected 
only by the final explosion, unless strong mass loss uncovers the former H-core (Wolf-Rayet
star), in which
case it is also ejected through the stellar wind; this 
happens in non-rotating stars with initial masses above $\sim$30 \ms ~at 
solar metallicity. In general, the more massive the star, the larger
the fraction of hydrostatically produced \Al in interstellar space.

The amount of ejected \Al depends on several factors: The specific criterion
adopted for modelling convection (i.e. ''Schwarzschild`` vs. ''Ledoux``) 
determines the size of
the convective stellar core and of the various burning shells. The reaction 
rate of $^{25}$Mg(p,$\gamma$)\Al is uncertain by a factor of $\sim$2 in
the relevant energy range (see e.g. the regularly updated 
NACRE reaction rate compilation\footnote{Web site at: {\tt http://pntpm.ulb.ac.be/nacre.htm}}). 
The still
poorly known rate of the $^{12}$C($\alpha,\gamma$) reaction determines
the amount of  $^{12}$C left in the core at He-exhaustion and, therefore, 
the amount of $^{20}$Ne produced through C-burning which, ultimately, 
determines the amount of $^{25}$Mg (a product of Ne-burning) available for
\Al production. Rotation has recently been introduced in massive star models.
It induces diffusion of species from the  core to the envelope and reduces the
minimum initial mass for a star to become WR \cite{pala04}. 
Finally, the abundant neutrinos from the collapsed stellar core may produce
additional \Al by spallating \Mg nuclei in the Ne-O layers, the exact yield
depending on the poorly known average neutrino energy 
\cite{woos95}. 

For various reasons, there has been no self-consistent 
model of massive star evolution including all the ingredients concerning
\Al production. Models evolved up to the SN explosion did not
include mass loss \cite[e.g.]{woos95,chief03,thie96} 
and thus underestimated the hydrostatically
produced part of \Al (because that amount of \Al decays inside the star
while waiting for the explosion, whereas it is rapidly brought to the
surface and ejected in the case of models with mass loss). On the other
hand, models including mass loss (and other ingredients, like rotation)
until recently did not not follow the evolution until the final explosion and thus completely miss
the explosive part of the yield of \Al. Finally, only the Santa-Cruz group of
theoreticians \cite{woos95,raus02} 
has included neutrino-induced nucleosynthesis in its model studies.

For all those reasons and uncertainties, it has been
rather difficult up to now either to compare the yields of various groups or
to make a self-consistent evaluation of the total amount of \Al expected
to be ejected from a population of massive stars, 
covering the full mass range of 10 up to 100~M$_\odot$.
Much attention has been paid to the comparison of 
hydrostatically-produced \Al as expelled through the winds of the most massive stars (WR),
and the one produced (both hydrostatically and explosively)
by somewhat less massive stars (below 30 \ms), 
for which mass loss is unimportant. Although such a comparison may appear 
futile, since it introduces an artificial separation between various
stellar mass ranges, it played a pivotal role in the development of the
whole field of the nucleosynthesis studies of $^{26}$Al: indeed, it pushed
theoreticians to continuous refinement of their models and
a much more thorough exploration of the various factors and uncertainties
affecting the \Al yields. 

The case of $^{60}$Fe, another long-lived isotope, again
exemplifies such an approach to modeling of nucleosynthesis. 
\Fe is produced in the same
zones as \Al (Ne-O zone), both hydrostatically and
explosively, by successive n-captures on $^{58}$Fe and  $^{59}$Fe.
It is also produced in the base of the He-shell, by some mild r-process
during the explosion \cite{woos95} (see Fig. 1). Uncertainties on its yield
are thus related to the cross-section of n-capture on unstable $^{59}$Fe 
and, of course, on the convection criterion employed. 
Contrary to the case of \Al, \Fe is expected to
be ejected only by the SN explosion and not by the stellar wind, since it is
buried so deeply (in the Ne-O shell). For this reason, it has been suggested 
that detection of \Fe in the Galaxy would help to decide whether WR stars
or core-collapse SN are the major sources of observed \Al (see Sec. 3.3).

This unfortunate artificial division
between massive stars evolving with $\sim$constant mass and stars with
mass loss is reaching an end now, since the
first results of complete models (now including mass loss, but still no rotation, and 
reaching out to the SN stage) have been recently reported \cite {limo04}; in those
models, explosive ejection of \Al always appears to dominate the stellar-wind ejected
$^{26}$Al, even for the most massive stars (so much, in fact, that
overproduction of \Al might become an issue...).

The other major radioactivities from massive stars (\Ti, \Ni, $^{57}$Co) are produced
very near to the collapsed Fe-core of the star, by explosive Si-burning at 
temperatures T$>$4~10$^9$~K, or in the regime of 
Nuclear Statistical Equilibrium (NSE, T$>$5~10$^9$~K). The production of
\Ti also requires conditions of relatively low density, so that the
alpha-particles which are abundantly produced during NSE do not quickly combine to
form $^{12}$C through the 3$\alpha$ reaction 
(which is very sensitive to density, being a 3-body reaction); a large fraction
of mass in free $\alpha$-particles after termination of NSE (the so-called $\alpha$-{\it rich 
freeze-out}) is the necessary condition for significant production of \Ti.
\Ni is mostly produced in the NSE phase, hence not sensitive to 
 $\alpha$-rich freeze-out, while \Ci is mildly so.
All three isotopes are produced in regions of small  neutron 
excess (electron mole fractions
$Y_e>$0.498) and are very sensitive to the, still very poorly
known, explosion mechanism (e.g. \cite {jank04}); 
in particular, they are sensitive to the position of the
``mass-cut'', the fiducial surface separating the supernova ejecta from
the material that falls back to the compact object after the
passage of the reverse shock (see Fig. 1).

In the case of SN1987A, a supernova which occurred in the nearby Large Magellanic Cloud
from a progenitor star with mass $\sim$18-20 \ms,
the extrapolation of the optical light curve (powered
by \Co radioactivity) to the origin of the explosion indicates a production of 
0.07~\ms ~of \Ni \cite{arne89}; this is sometimes taken as a ``canonical'' yield
of \Ni from ccSNe and often used in calculations of
galactic chemical evolution. 
However, optical observations (albeit
with large uncertainties) show that ccSNe display
a wide range of \Ni values, correlated with the energy of the explosion
\cite{hamu03}.
The late lightcurve of SN1987A also 
constrains the amounts of other radioactivities and, in particular
that of \Ti (see Sec. 3.1); however, only direct observations of the characteristic
gamma-ray lines can confirm theoretical predictions.
Care must be taken to acknowledge the diversity of events - ccSNe are not a homogeneous 
class of events.

NSE conditions are also encountered in the innermost regions of thermonuclearly
exploding supernovae of type Ia. 
Uncertainties here are related to the progenitor system, to the way it
reaches the Chandrasekhar mass and how it collapses, to the ignition density, and to the
propagation of the nuclear flame (which determine the amount of electron
captures and the degree of neutronisation of the ejecta, e.g. \cite {hill04}). 
In thermonuclear supernova explosions, the evolution of flame speed 
determines the
nucleosynthesis and in particular the total amount of \Ni produced.  
Anyone of the common SNe~Ia scenarios (sub-Chandrasekhar, deflagration,
delayed detonations, and pulsating delayed detonation models) 
seems able to produce a wide variety of $^{56}$Ni masses, 
ranging from $\simeq$~0.1 to 1~\Msol\  \cite{nomo84,hoef96,hoef02}.
Generically, SNIa release about ten times as much \Ni than 
 ccSNe. SNIa have lower envelope
masses and higher expansion velocities than ccSNe ($\sim$0.5 \ms ~against 
several \ms, and  $\sim$2~10$^4$~km~s$^{-1}$ 
against $\sim$5~10$^3$~km~s$^{-1}$), reflecting their presumed origin from a
star which has previously lost its envelope.
For those reasons they are expected to become transparent to $\gamma$-rays much
earlier and to be much brighter $\gamma$-ray sources than ccSNe. 
The expected  $\gamma$-ray line fluxes from the \Co \ decay, combined 
to sensitivities of  present-day $\gamma$-ray instruments, 
limit observations of SNIa to rare events within about 15 Mpc 
(i.e. one every few years). 
Because of the lacking envelope, SNIa are expected to also be much 
more important sources
of positrons than SNII (and, in fact, than any other known 
source), although the exact amount of escaping e$^+$
depends on poorly understood factors, like the intensity and configuration
of the magnetic field in the exploded star (see Sec. 3.4). 

In the light of the frequencies of occurence of ccSNe and SNIa 
in external galaxies
\cite[e.g.]{Mannucci04} it is a sign of fortune that
SN1987A is the first, 
and, up to now, only one supernova which is clearly detected 
in the light of its radioactivity $\gamma$-rays.
We still lack a complete picture on how radioactive energy is deposited 
inside supernovae into other forms of energy (radiating, or kinetic). 
In spite of the important $\gamma$-ray signals from SN1987A and Cas~A, 
those are just two observed events, which may not sample an average ccSN, especially
since core collapse by itself seems not so tightly regulated to produce 
a rather homogeneous event class such as SNIa are. Nevertheless, the physics of such 
radioactive-energy deposits probably will be revealed through observations
of many more such events, probably then most directly in 
more signatures at $\gamma$-ray energies.

	
\section{SPECIFIC ISOTOPES AND SOURCES}
   

\subsection{Fe Group Nuclei and $^{44}$Ti from  supernovae}

{\bf Thermonuclear Supernovae:}
SN1991T occurred at a distance of 13 Mpc, and was a peculiar and exceptionally-bright
supernova of the Ia type. Its Co decay lines were marginally detected 
(at a significance level of 
3-5 $\sigma$) with the COMPTEL telescope aboard the Compton 
GRO \cite{morr97}. The mean flux in the two $^{56}$Co decay lines
at 847 and 1238 keV was found to be 
1.17$\pm0.32\pm0.35$~10$^{-4}$~ph~cm$^{-2}$s$^{-1}$
(uncertainties from statistics and systematics, respectively)
\cite{morr02}. When converted to \Ni mass, 
a best estimate of 1.5~\Msol\ is obtained,
with a lower limit of 0.65~\Msol\ which accounts for uncertainties
in measurement and distance
to the supernova \cite{morr02}. This is consistent with other estimates of
SN1991T's \Ni mass, for this peculiarly-bright event \cite{leib00}.
 
SN1998bu provided a second
and seemingly better opportunity for the instruments on Compton GRO, since it
exploded at a distance of 11.6~Mpc \cite{hjor97}. 
Yet, no gamma-rays from \Co \ decay
were observed, despite a total exposure of fourteen weeks 
(compared to only  two weeks in  the case of SN1991T) \cite{geor02}. 
The COMPTEL limit for the 1238~keV line of \Co ~ 
(2.3 10$^{-5}$~photons~cm$^{-2}$~s$^{-1}$) 
constrains the visible $^{56}$Ni mass to 
below 0.35 \Msol\ if the supernova is assumed to be completely transparent
to gamma-rays at the time of observations. This is probably not the case, 
however, and a large fraction of the $\gamma$-ray energy should be deposited 
in the supernova during this time window. Observations in other wavelengths suggest
that  0.7-0.8 \Msol\ of $^{56}$Ni were produced in SN1998bu \cite{leib00,sollerman04}.
Detailed Monte Carlo energy transport calculations show then
that COMPTEL should have seen $^{56}$Co  $\gamma$-rays, 
at least in the framework of  the ``brightest'' of those models
(those that turn more rapidly from deflagration into detonation (W7DT) 
or partially-produce radioactivity in their outer
ejecta (HECD) \cite{geor02}). 

\begin{figure}
\centering\
\includegraphics[width=.6\textwidth]{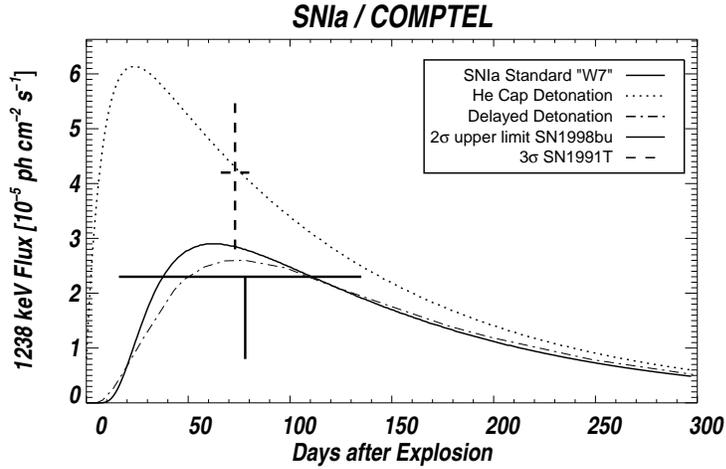}
 \caption{Gamma-ray light curves from SNIa models, as compared to the
 measurements from SN1991T and SN1998bu (1238 keV line from $^{56}$Co decay).}
  \label{snia_gammas}
\end{figure}

From these two events, gamma-ray results on thermonuclear supernovae remain
puzzling (see Fig. \ref{snia_gammas}). Simulations show that even when 
$\gamma$-ray spectra  
can be measured with high accuracy, probably an event distance 
well below 10~Mpc is needed for a convincing classification of the 
explosion type \cite{iser04}. 

{\bf Core Collapse Supernovae:}
In the case of ccSNe, the characteristic $^{56}$Co 
decay $\gamma$-ray lines have been unambiguously observed in the 
relatively nearby (distance $\sim$55 kpc) SN1987A,
both with the low-resolution NaI detector aboard the Solar Maximum Mission (SMM)
\cite{matz88}, as with a balloon-borne Ge detector\cite{tuel90}; the latter 
showed the line to be slightly red-shifted and broadened, a fact that has not 
received a convincing explanation yet (see e.g. \cite{gran93}). 
A few years later, the OSSE instrument aboard the Compton GRO 
detected the 122~keV line from the decay of $^{57}$Co \cite{kurf92}, and 
probed further the physical conditions in the innermost exploding layers of
the supernova \cite{clay92}.

 Another important diagnostics of core-collapse supernovae is provided by \Ti, the
   parent isotope of the stable and abundant in nature  $^{44}$Ca \cite{woos73}.
   The most plausible cosmic environment for \Ti~production  is
   the $\alpha$-rich freeze-out from high-temperature burning near
   Nuclear Statistical Equilibrium, or from Silicon burning (e.g. \cite{arne96}).
   Both processes occur in the innermost layers of core-collapse supernovae, which
   are thought to synthesize substantial amounts of $^{44}$Ti, along with $\sim$1000
   times more \Ni (see Fig.~\ref{TivsNi}).  The discovery of \Ti decay products in 
   presolar grains \cite{nitt96}, and the modelling of the late 
   bolometric light curve    of  SN1987A with  energy input from 
   \Ti radioactivity \cite{fran02} 
   provide support to those ideas.  
  
   The lightcurve of SN1987A has been observed in unique  detail for more 
   than  15~years. After decay of the initial \Ni and \Co 
   it appears now powered by \Ti radioactivity. The amount of \Ti is estimated to
   1--2.~10$^{-4}$M$_{\odot}$, from recent modeling of
   radioactive energy deposition and photon transport in the SNR \cite{fran02,moti03}.
   From infrared observations, an even tighter upper limit of 1.1~10$^{-4}$M$_{\odot}$
   has been derived \cite{lund01}.
   Gamma-ray detection and proof of this interpretation is still 
   lacking; this would be most direct. Unfortunately, the derived
    amount of \Ti and the distance of SN1987A 
    result in a $\gamma$-ray line flux slightly below  INTEGRAL's sensitivity
    \cite{moti03a}.
  
\begin{figure}
\centering\
\includegraphics[width=.45\textwidth]{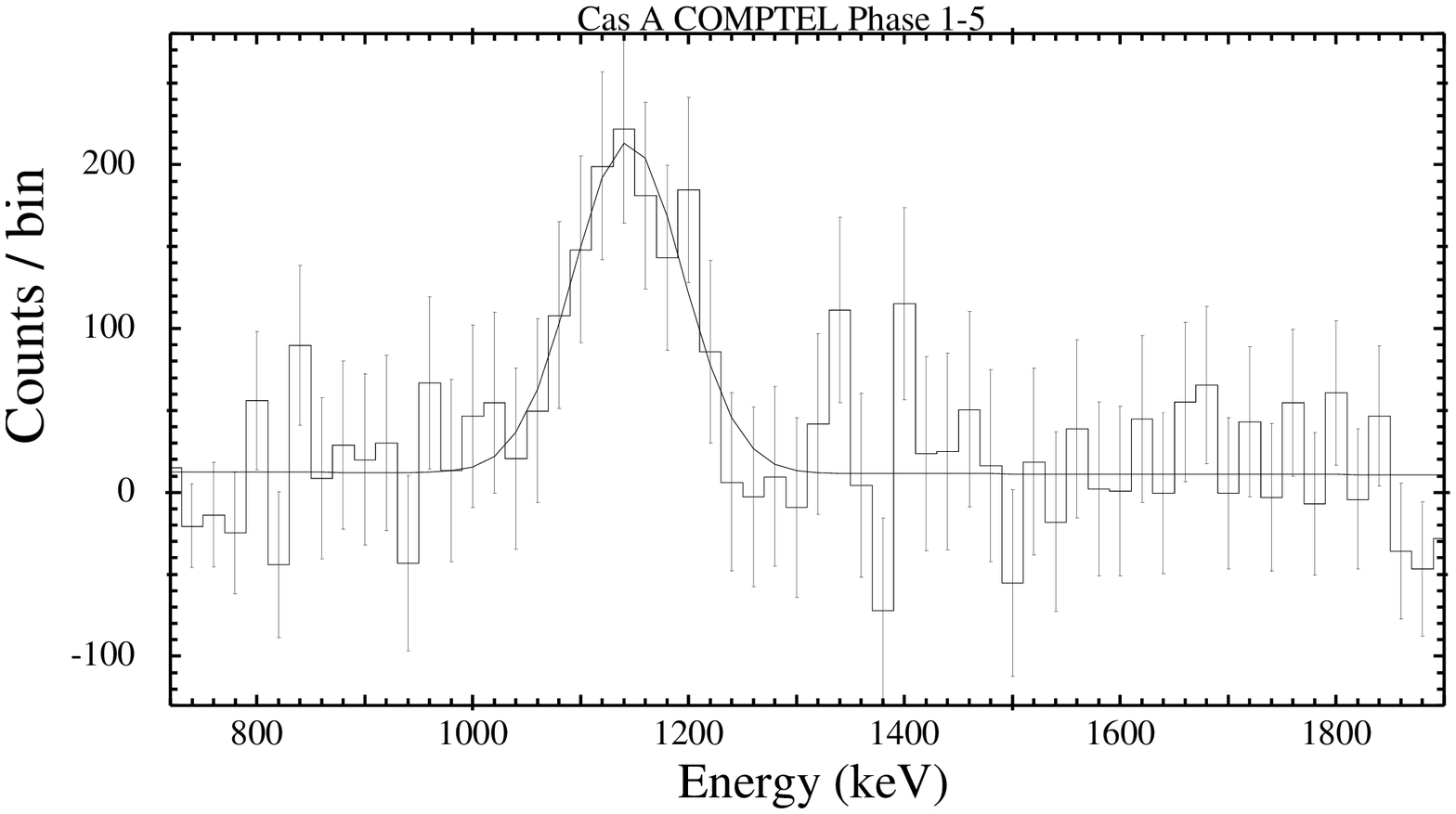}
\includegraphics[width=.45\textwidth]{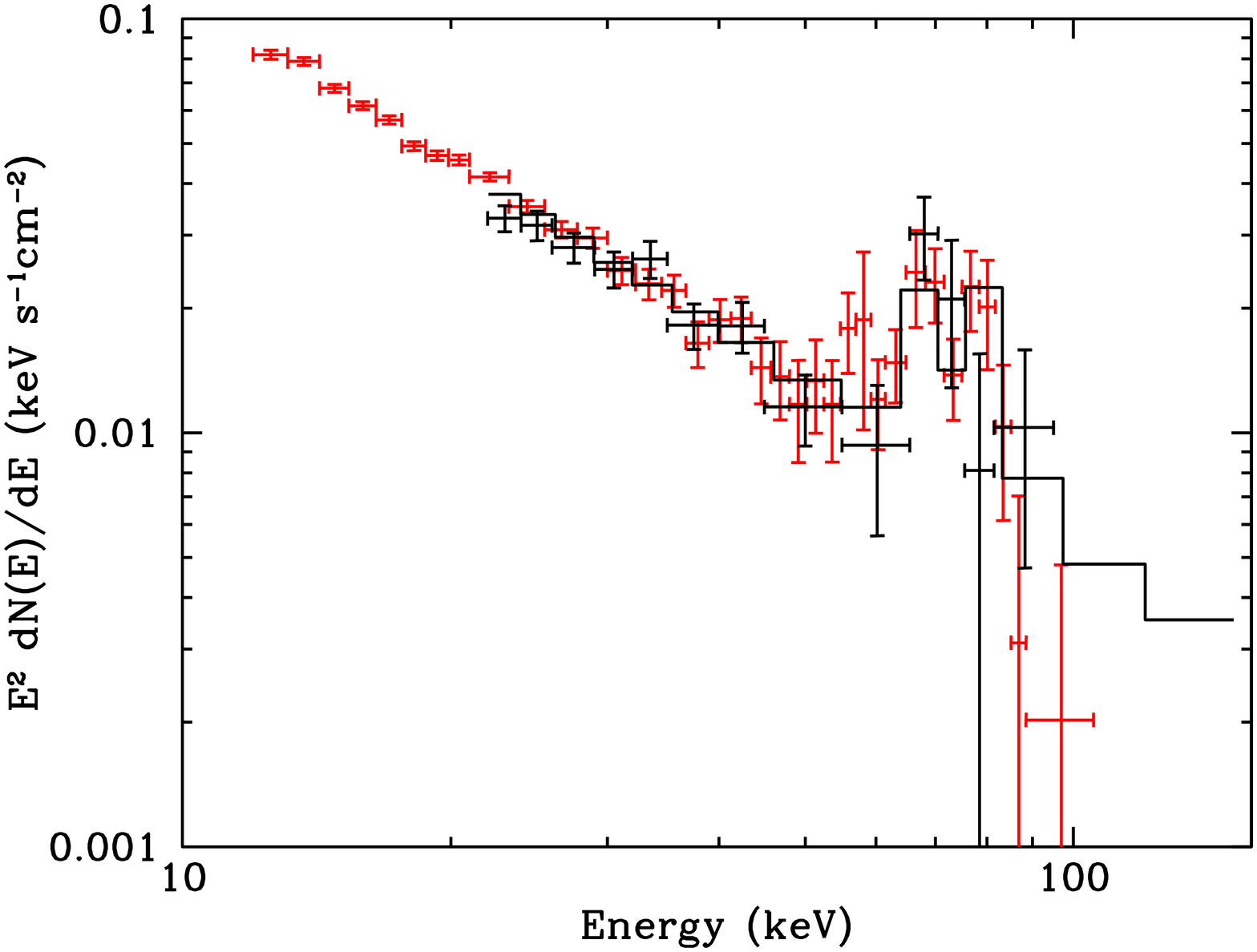}
 \caption{Cas A supernova remnant \Ti measurements from ({\it left}) the COMPTEL instrument 
 (1156 keV $\gamma$-ray line \cite{iyud97}) and
 from ({\it right}) the BeppoSAX ({\it red}) and INTEGRAL/IBIS ({\it black}) instruments 
 (68 and 78~keV $\gamma$-ray lines \cite{vink01,vink05}).}
  \label{casa_spectrum}
\end{figure}

\begin{figure}
\centering
\includegraphics[width=.7\textwidth]{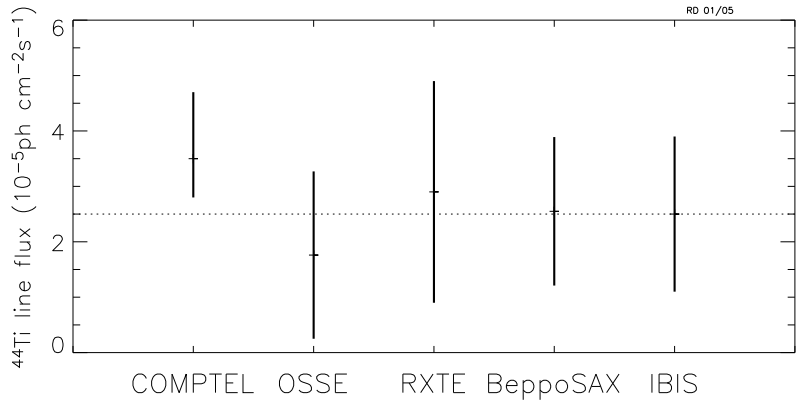}
\caption{
Intensity measurements 
from different experiments for \Ti decay emission
from Cas A \cite{scho00,the96,roth97,vink01}. (For COMPTEL and BeppoSAX,
systematic uncertainty from a possible underlying continuum has been
added quadratically to the statistical uncertainty). A flux value 
of 2.5~10$^{-5}$~ph~cm$^{-2}$s$^{-1}$ appears reasonable and corresponds
to 1.5~10$^{-4}$M$_{\odot}$) of $^{44}$Ti. 
 }
\label{CasA_fluxes}  
\end{figure}

\begin{figure}
\centering
\includegraphics[angle=-90,width=.8\textwidth]{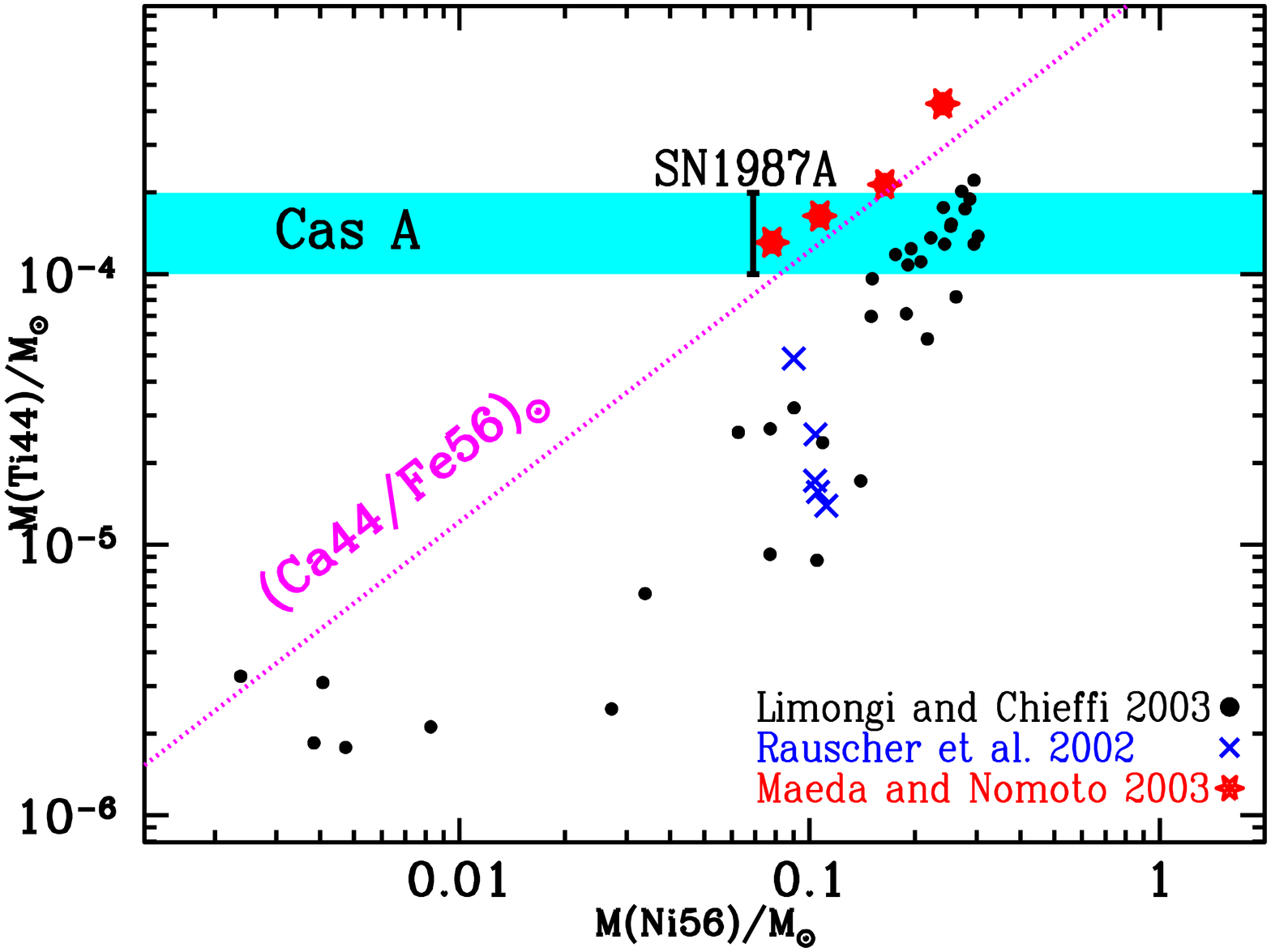}
\caption{
Yield of \Ti \ vs yield of \Ni, from models and observations.
Model results are from Limongi and Chieffi (2003, filled dots, with 
large variations in yields due to variations in both stellar mass - from
15 to 35 \ms \ - and
explosion energy), Rauscher et al. (2002, crosses, for stars in the
15 to 25 \ms \ range and explosion energies of 10$^{51}$ ergs) and
Maeda and Nomoto (2003, asterisks); the latter concern
axisymmetric explosions in
25 and 40 \ms \ stars, producing high \Ti/\Ni \ ratios. 
\Ti \ detected in Cas~A appears as
a horizontal shaded band (assuming that its decay rate has not been
affected by ionisation in the Cas~A remnant, otherwise its abundance should be
lower, according to Motizuki et al. 1999). 
The amount of \Ti \ in SN1987A is derived
from its late optical lightcurve (Motizuki and Kumagai 2003, see Fig. 4).
The diagonal dotted line indicates the solar ratio of the corresponding
stable isotopes ($^{44}$Ca/ $^{56}$Fe)$_{\odot}$(from \cite{prantzos_iws04}.  }
  \label{TivsNi}
\end{figure}

   The 1.156 MeV $\gamma$-rays following 
   \Ti decay have been detected in the 340-year old Galactic supernova remnant 
   Cas~A \cite{iyud94}, at a distance of $\sim$3.4 kpc  (Fig.~\ref{casa_spectrum}). 
   The analysis of the data cumulated by COMPTEL over the years shows that the
    detection is clearly significant ($>$5$\sigma$), although
   the initially reported flux was too high, and a value of
   3.4~10$^{-5}$~ph~cm$^{-2}$s$^{-1}$ has been assessed \cite{scho00}.  
   A few other attempts to confirm this
   discovery did not succeed due to strong instrumental backgrounds 
   (see Fig.~\ref{CasA_fluxes}). The BeppoSax instrument has obtained
   a significant and convincing measurement of the low energy lines 
   from the \Ti decay chain at 68 and 78~keV, respectively \cite{vink01}. 
   Their combined flux (at 3.4$\sigma$ significance) is quoted as 1.9 and 
   3.2~10$^{-5}$ ph~cm$^{-2}$s$^{-1}$, for two different assumptions about the
   underlying continuum. Varying in spectral
   shape between a simple power-law and a steepening bremsstrahlung spectrum,
   this continuum  constitutes a major systematic uncertainty of all these 
   measurements.
   The results from different instruments show overall uncertainties
   on the order of 30--50\%, so that a flux value of 
   2.5$\pm$1 10$^{-5}$ ph~cm$^{-2}$s$^{-1}$ 
   is suggested (Fig.~\ref{CasA_fluxes}).
   Note that \Ti decays through electron capture, which is considerably
   slowed down in
   ionised media.  Our poor knowledge of the ionization state of \Ti
   in the young supernova remnant adds further uncertainty to the derived 
   yield  \cite{moti99}; recent model studies obtain values similar to the
   case of SN1987A, i.e. in the range 1-2~10$^{-4}~$\ms (Fig.~\ref{TivsNi}).


The ejection of \Ni and \Ti from the inner regions of a ccSNe
depends critically on the kinetic energy and the mechanism of the explosion. 
 A  substantial amount of the elements synthesized in the innermost layers 
 may fall back onto the compact remnant.
As the supernova explosion is not yet understood in sufficient detail
for quantitative modeling, current studies of supernova nucleosynthesis adopt 
parametric descriptions: either the supernova explosion energy is injected 
as thermal energy at the inner boundary, or a mechanical "piston"
is assumed to impart the kinetic energy of the explosion to the ejecta \cite{thie96,woos95}. 
Due to these uncertainties, model yields suffer from large uncertainties, in particular
for \Ti which is mostly produced near or even inside the mass cut (see Fig. \ref{radioact_profiles}). 
Kinetic energies of the explosion have been estimated as 1.2 and 2~10$^{51}$~erg for
SN1987A and Cas~A, respectively \cite{lami03,utro04}; taken 
at face value, those numbers suggest that more \Ti and \Ni
should have been ejected in Cas~A than in SN1987A. Since the peak SN 
luminosity is proportional to the \Ni~yield, it is then rather 
surprising that the explosion of Cas~A at only 3.4~kpc distance 
was not reported by contemporaneous 
observers around year 1671 \cite{thor01}. Occultation of the Cas~A supernova by 
circumstellar dust, which was 
then destroyed by the supernova blast wave, was suggested \cite{schmi95}.
Baring this and other more exotic
 solutions,  one concludes that the ejected  
\Ni amount was not overly high and rather  less than the 0.07~M$_{\odot}$ 
estimated for SN1987A. Still, Cas~A must have produced a rather high
\Ti \ yield, in view of the gamma-ray observations.
This is at odds with one-dimensional models for ccSN nucleosynthesis 
(see Fig.~\ref{radioact_profiles}).

Another approach to estimate \Ti yields for the average ccSN
is based on standard cosmic abundances, 
which should be reproduced by models of galactic chemical evolution
that use the same model yields as those probed by gamma-rays.
For that purpose, it is noted that about half of solar $^{44}$Ca is produced
as \Ti \ in ccSNe, which also produce about half of solar $^{56}$Fe as \Ni \
(the other half coming from thermonuclear SN).
It is expected then that in ccSNe the ratio of the unstable parent
nuclei \Ti/\Ni \ should match the solar ratio of the corresponding
stable daughter nuclei ($^{44}$Ca/$^{56}$Fe)$_{\odot}\sim$10$^{-3}$. However,
recent supernova nucleosynthesis calculations obtain $^{44}$Ti/$^{56}$Ni 
values $\sim$3~times lower than that (Fig.~\ref{TivsNi}). Taken at face value, this result 
implies that such explosions cannot produce the solar $^{44}$Ca, otherwise 
$^{56}$Fe would be overproduced  \cite{timm96}.

Obviously, higher $^{44}$Ti/$^{56}$Ni ratios are required from stellar
models in order to
satisfy the requirements of galactic chemical evolution, and such high
ratios are also required to explain the high value of the \Ti \ yields
in SN1987A and Cas~A.


An exciting possibility to
simultaneously solve those problems arises from the potentially important role
of asymmetric explosions -- this could provide the necessary boost in \Ti yields.
Multi-dimensional models of supernova nucleosynthesis
are difficult and computationally challenging at present. Still,
prelimary parametric calculations indicate that \Ti production can 
be increased substantially through such asymmetries, ejecting more mass in 
polar regions of a rotating star during the explosion \cite{naga97,maed03} 
(see Fig. \ref{TivsNi}).  
It remains to be shown that such models can satisfy the constraints of the
Standard Abundance Distribution for other elements, such as e.g. the
isotope ratios of Ni.

Both in Cas A and SN1987A, preferred directions are evident.
The bright rings seen in SN1987A's remnant suggest that rotation must 
have been high in the pre-supernova star to produce such torus-like dense gas
during the pre-supernova wind phase. Further evidence is provided by polarisation
measurements \cite{wang02}. 
In the case of Cas~A, fast-moving knots of ejecta within
a standard-expanding remnant had been reported early on \cite{fese87}.
Early indications for a jet from optical data \cite{fese96} have recently been
confirmed with Chandra \cite{hwan04}, where silicon-rich jet structures
in opposite directions have been imaged in characteristic X-rays.
Therefore, the diagnosis of the dynamics of \Ti ejection in Cas~A,
expected from  deep INTEGRAL observations, promises interesting insights to
this exciting problem.


Statistical constraints on the \Ti yields of supernovae can be obtained
through surveys of the Milky Way and search for \Ti emission from young
SN remnants (like Cas~A, or younger); indeed, taking into account the estimated SN rate
in the Galaxy (about 2-3 core collapse SN per century), one expects that
a few of them should be detectable with present day instruments.
COMPTEL's survey in the 1.156 MeV
band of \Ti emission is still the most complete one \cite{dupr97,iyud99}.
Apart from the clearly detected Cas A, a couple of 
candidates at low significance have been discussed
(most prominently GRO~J0852-4642 in the Vela region \cite{iyud98},
 and a weak signal from the Per OB2 association \cite{dupr97}).
 First INTEGRAL results cast some doubts on the COMPTEL source of \Ti in
 the Vela region \cite{kien04}, although its spatial coincidence with a newly-discovered and 
 rather nearby ($\leq$~1~kpc) X-ray SNR has led to interesting speculations 
 about a very nearby supernova event \cite{asch99}. 
Apparently, no bright young \Ti emitting supernova remnants are found in the
inner regions of the Galaxy. Although marginally consistent with the 
uncertainties of such low-number statistics, it appears as if the 
ejection of ``average'' amounts of \Ti is not common in core-collapse 
supernovae. Monte Carlo simulations   
and their normalizations to COMPTEL data and to
historic records of supernova observations of the last millenium
\cite{the00} suggest that a rather rare supernova type with high
\Ti yield is favoured by the absence of a \Ti signal from the inner Galaxy.
Similar conclusions are reached by preliminary INTEGRAL data analysis
\cite{rena04}.


\subsection{Positrons, $^{22}$Na and $^{7}$Be from Novae}

Nova explosions are the result of accretion of a critical mass of H-rich material
on the surface of a white dwarf in a close binary system, which leads to
a thermonuclear runaway.  Explosive H-burning via the hot CNO-cycle 
produces several radioactive species, which decay emitting gamma-rays and positrons.

Positrons emitted from the short-lived isotopes $^{13}$N and $^{18}$F
annihilate with electrons inside the nova envelope and produce a prompt $\gamma$
ray emission, which appears very early (before optical maximum), lasts for 
$\sim$2 days and consists of a 511 keV line and a continuum between 20 and 511 keV.
The decay of the long lived  \Be and \Na produce  $\gamma$-ray lines
at 478 keV and 1275 keV, respectively (see Table 1). \Be is produced by
$^4$He+$^3$He in classical CO nova (i.e. with a white dwarf composed of C and O)
while \Na is produced in the hot Ne-Na cycle which occurs in ONeMg novae
(with a heavier white dwarf, resulting from the evolution of 6-9 \ms \ stars).
Thus, the detection of the caracteristic $\gamma$-ray lines of \Be
and \Na would allow to
unambiguously identify the composition of
the progenitor white dwarf of the nova system. On the
other hand a (serendepitous!)
detection of the early 511 keV emission would give invaluable
information on the explosion itself and the degree of mixing of 
the H-burning  products in the envelope. Finally,
ONeMg nova may also produce interesting amounts of \Al; however, the
COMPTEL sky map of 1.8 MeV emission in the Milky Way strongly argues
for massive stars as the main source of galactic \Al (see next section).

Current nova models (with 1-D hyrodynamics)
are quite successful in explaining several observed
propreties, like light curves, abundances and velocities of the ejecta
\cite{hern99,starr98}.
However, they also suffer from uncertainties related
in particular to the amount of mass ejected, which is systematically found
to be lower than observed by factors $\sim$10; indeed, nova models
predict ejected masses of $\sim$10$^{-5}$ \ms, while
observations suggest values closer to $\sim$10$^{-4}$~\ms. This uncertainty
is  reflected in the predicted intensities of the 478~keV and 1275~keV line
and to the maximum distances for a nova explosion to be detectable by a 
given instrument.

Sky surveys of the 478~keV line (with the GRS instrument aboard the
SMM satellite \cite{harr91}
and of the 1275~keV line (with
GRS on SMM and with COMPTEL aboard CGRO \cite{iyud99} 
have
provided only upper limits to the corresponding fluxes. These are
fully compatible with current nova models \cite{jose99} 
which predict fluxes of 
2~10$^{-6}$~ph~cm$^{-2}$s$^{-1}$ and  2~10$^{-5}~$ph~cm$^{-2}$s$^{-1}$
for the \Be and \Na lines, respectively, for novae at a distance
of 1 kpc. In view of  the sensitivity of  
the SPI instrument aboard INTEGRAL, the maximum distances for a
nova to be  detectable are $\sim$0.5~kpc and $\sim$2~kpc, for
the \Be and \Na lines, respectively. Taking into account 
the observed frequency of novae closer than those limits, it is
expected that one such explosion will be detected by INTEGRAL
in the next few years.  


\subsection{$^{26}$Al and $^{60}$Fe: Large-Scale Galactic Nucleosynthesis}

\begin{figure}
\centering
\includegraphics[width=0.8\textwidth]{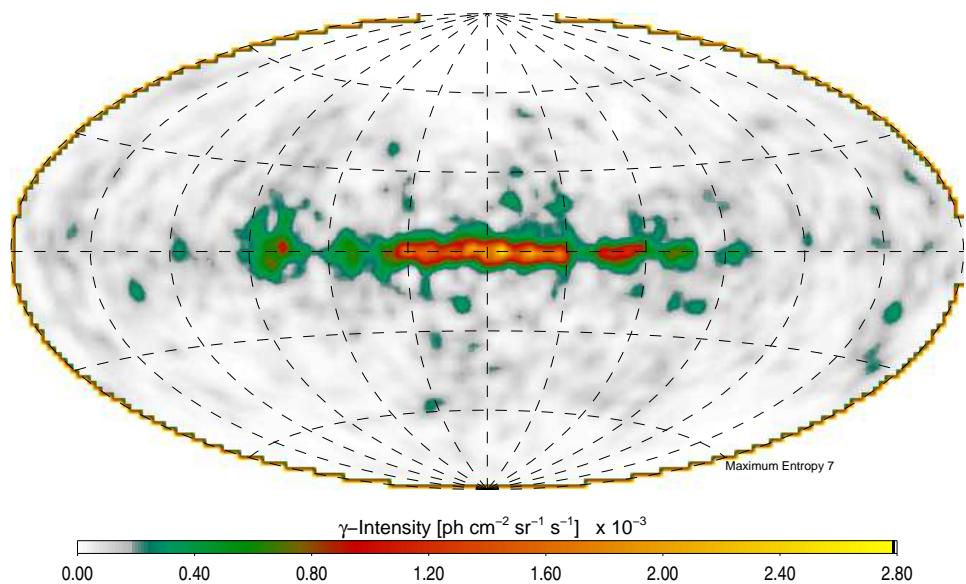}
\caption{All-sky image of \Al gamma-ray emission at 1809~keV as derived from
COMPTEL's 9-year survey \cite{plue01}.  
}
\label{almap}
\end{figure}
  
\begin{figure}
\centering
\includegraphics[width=7.5cm]{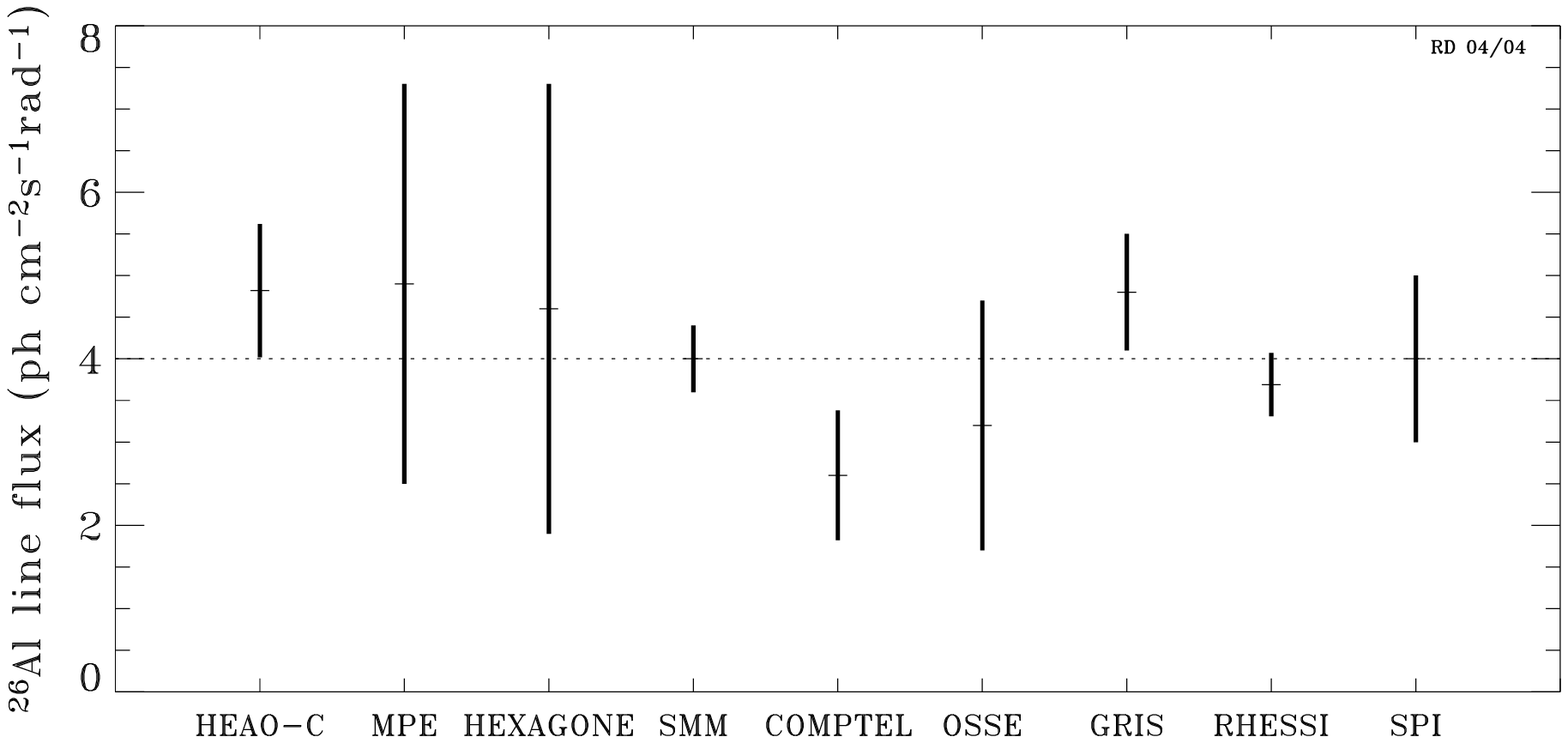}
\includegraphics[width=7.5cm]{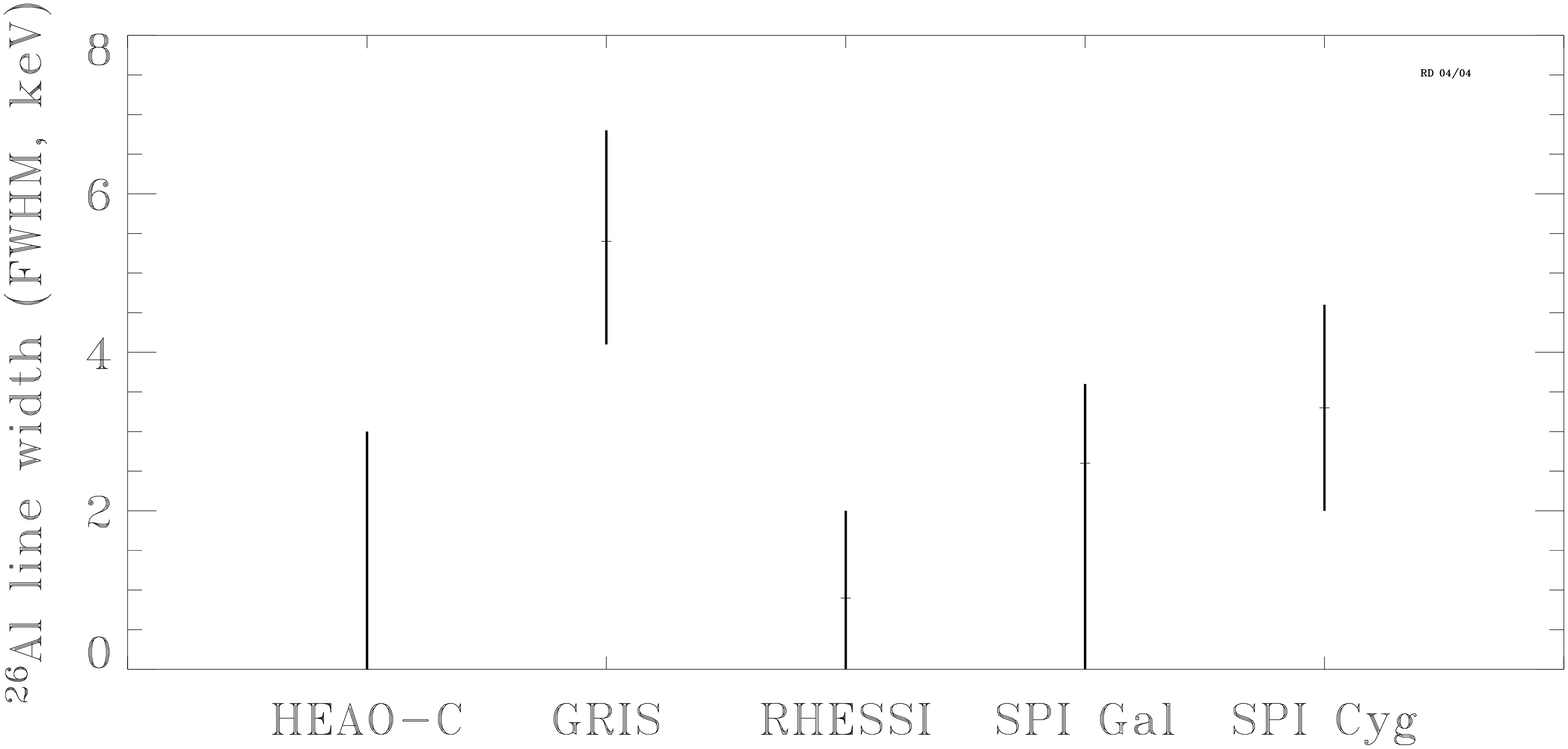}
\caption{Intensity measurements (left) and line width measurements (right)
from different experiments for \Al emission
from the inner Galaxy, and for Cygnus (rightmost datapoint). }
\label{diehl_26Al_experiments}
\end{figure}

\begin{figure}
\centering
\includegraphics[width=7cm]{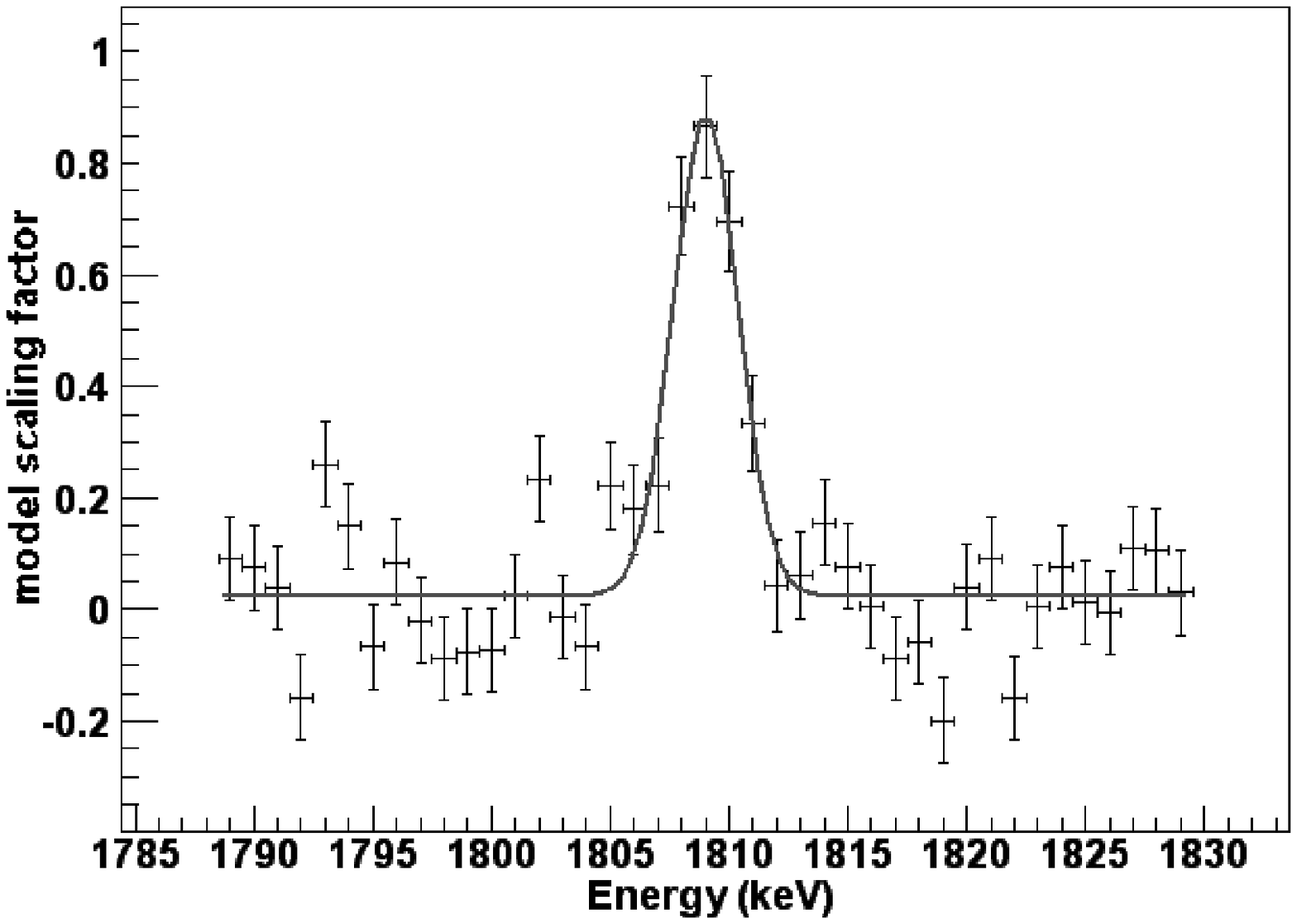}
\includegraphics[width=7.2cm]{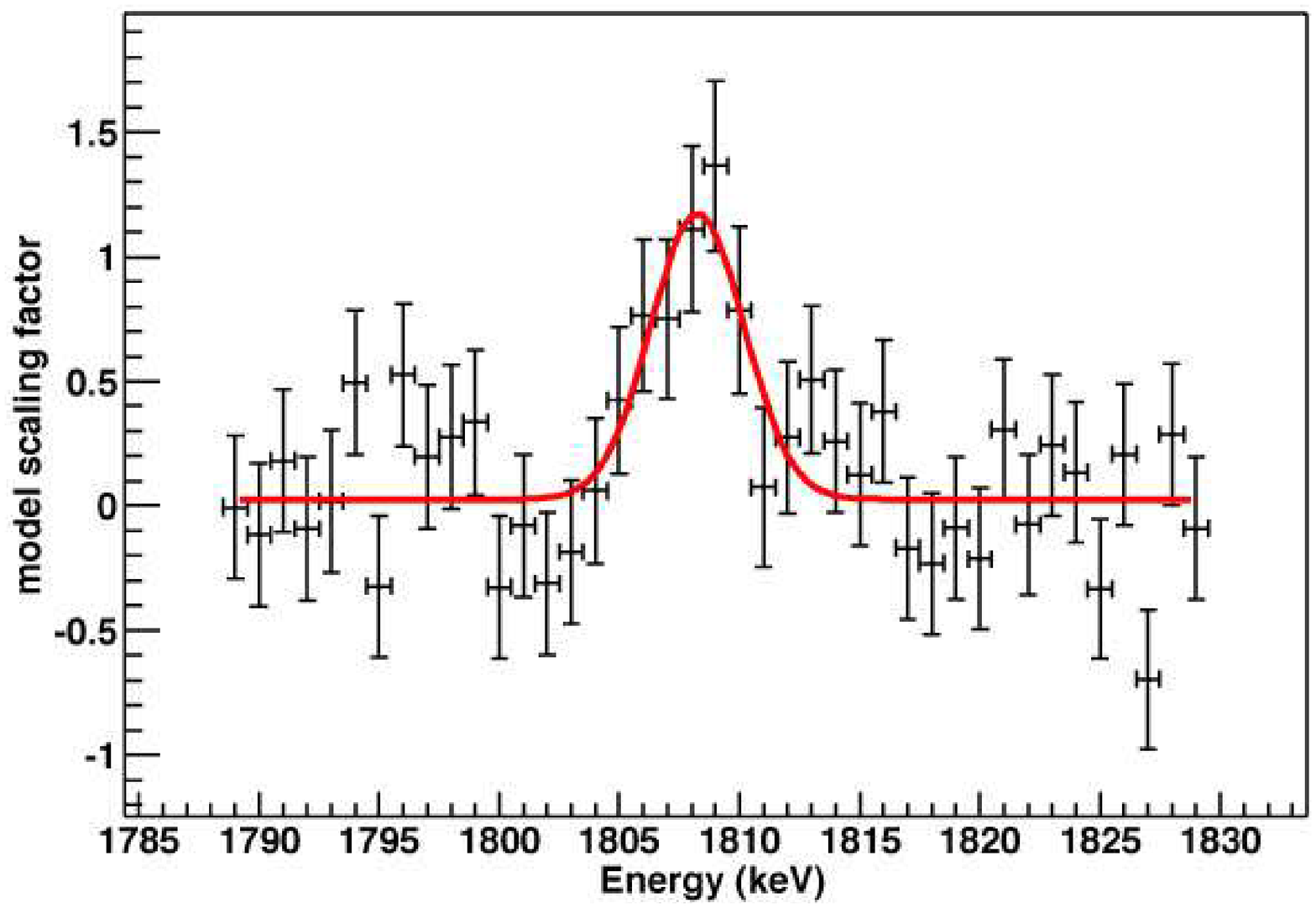}
\caption{SPI \Al line spectra for the inner Galaxy (left) and the
Cygnus region (right). The spectra are derived by fitting skymap intensities
per energy bin. Exposures are \about~4Msec each, from the Core Program
inner-Galaxy survey, and from 
Cygnus region calibrations and Open-Program data.  
}
\label{fig_spec_obsfit}
\end{figure}

The COMPTEL sky survey over nine years had convincingly confirmed
the spectacular discovery of live \Al in the interstellar medium
by the HEAO-C instrument \cite{maho82}. With its imaging capability, COMPTEL 
had mapped structured \Al emission, extended along the plane of the Galaxy 
\cite{plue01,knoe_img99,ober97,dieh95} (see Fig.~\ref{almap}), in broad agreement
with earlier expectations \cite{pran91,pran93}.
Models of \Al emission from the Galaxy and specific localized source regions
have been based on knowledge about the massive-star populations, and suggest that
such stars indeed dominate \Al production in the Galaxy 
\cite{pran96,knoe_mod99,knoe_phd99}. 
Galactic rotation and dynamics of the \Al gas ejected into the 
interstellar medium are expected to leave
characteristic imprints on the \Al line shape  \cite{kret03}. 
The GRIS balloon experiment carried high-resolution Ge detectors, and had obtained 
a significantly-broadened line \cite{naya96}, which translates into a 
kinematic (Doppler) broadening of 540~km~s$^{-1}$. 
 Considering the $1.04~10^6$~y decay time of $^{26}$Al,
such a large line width is hard to understand, and requires
either kpc-sized cavities around \Al sources or major \Al 
condensations on grains \cite{chen97,stur99}. 
Alternative measurements of the \Al line 
shape are then of great interest to settle this important issue.

Current results on large-scale
Galactic \Al line flux and width measurements are summarized 
in Fig. \ref{diehl_26Al_experiments}.
Precision follow-up measurements of 1808.7 keV emission from 
Galactic \Al have been one 
of the main science goals of the INTEGRAL mission \cite{wink03i}.
From INTEGRAL/SPI spectral analysis of a subset of the first-year's inner-Galaxy deep
exposure (``GCDE''), \Al emission was clearly detected  (Fig. \ref{fig_spec_obsfit} left) 
\cite{dieh03,dieh04} at a significance level of 5--7$\sigma$ (through fitting 
of adopted models for the \Al 
skymap to all SPI event types over an energy range $\Delta E\sim$80 keV
around the \Al line).  The line width was found consistent with SPI's instrumental
resolution of 3~keV (FWHM). These early SPI results  are in agreement
with RHESSI's recent findings \cite{smit03} and do not confirm
the broad \Al \ line reported by GRIS
(Fig. \ref{diehl_26Al_experiments}).
On the other hand, the first spectrum generated from SPI data for the Cygnus region 
(see Fig. \ref{fig_spec_obsfit} right for single-detector events \cite{knoe04c}) suggests
that the line may be moderately broadened in this region. This
may be caused by locally-increased interstellar turbulence from the particularly
young stellar associations of Cygnus.

The detailed mapping of the Galactic distribution of \Al, obtainable through
a determination of the distances to the ``hot-spots'' 
is one of the main
long-term objectives of INTEGRAL \cite{kret03}, since it will provide the most
accurate picture of recent star formation in the Milky Way 
\cite{dieh04}.
On the other hand, the study of individual 
``hot-spots'' indicated on the COMPTEL map bears 
on our understanding of the evolution of young stellar associations 
(in the cases of Cygnus, Carina 
and  Centaurus-Circinus) and even individual stars (in the case of Vela).

\begin{figure}[ht]
\centering   
\includegraphics[width=9cm]{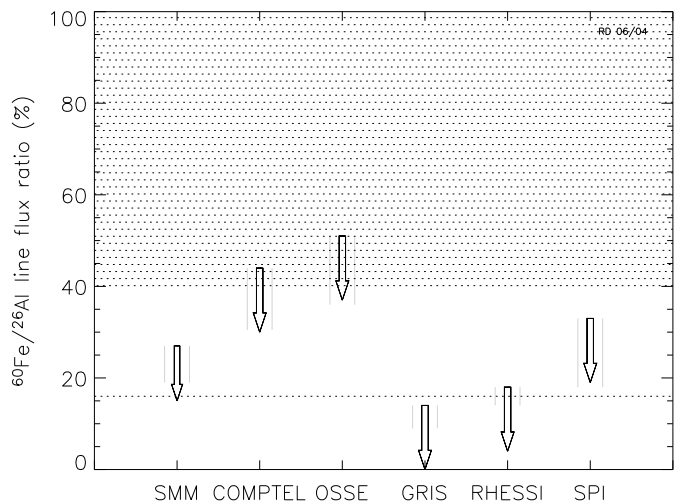}
\caption{Limits on the \Fe to \Al gamma-ray brightness ratio,
from several experiments fall below current expectations: 
An often-cited theoretical value from the 
Santa Cruz group \cite{timm96} is indicated as dotted
line, the dotted area marks the regime suggested by 
recently-updated models, from \cite{pran04}.
	 }
      \label{diehl_fe60limits}
\end{figure}

The fact that \Fe has not been clearly seen from the same source regions
appears surprising \cite{pran04}, given that
massive stars are expected to eject both \Al \ and \Fe in 
substantial amounts \cite{timm96,limo04}. RHESSI reported a
marginal signal (2.6 $\sigma$ for the combined \Fe \ lines 
at 1.173 and 1.332 MeV) \cite{smit04} from the inner Galaxy,
at the 10\%-level of \Al brightness;
SPI aboard INTEGRAL obtains a similarly low value around 10\%, 
also at the 3$\sigma$-level \cite{harr05,knoe04f}. Obviously, 
$^{60}$Fe $\gamma$-ray intensity from the inner Galaxy remains substantially 
below its \Al brightness. The situation becomes rather
uncomfortable, since current nucleosynthesis models in
massive stars suggest a large production of \Fe \
\cite{raus02,limo04}, substantially larger than older calculations
(see \cite{pran04} and references therein).
This is mainly due to increased neutron capture cross sections for Fe isotopes, 
and a reduced $^{59}$Fe $\beta$-decay rate. 
The expected gamma-ray line flux ratio of \Fe/\Al \ falls  
between 40 and 120\%, 
depending on how the interpolation between the few calculated stellar-mass
gridpoints is done, and which of the models are taken as baseline; in comparison,
the IMF choice appears uncritical. In any case, those
revised expectations clearly lie above the experimental limits for Galactic 
\Fe emission (Fig.~\ref{diehl_fe60limits}). 
Obviously, observations do not support predictions of current stellar
nucleosynthesis (or, alternatively, they suggest that ccSNe are not the
dominant sources of Galactic \Al).
The answer to this interesting nucleosynthesis puzzle may be related
to nuclear-physics issues, probably concerning the (uncertain) 
neutron capture reactions on unstable $^{59}$Fe.

\subsection{Positron Annihilation}

Ever since Anderson's discovery of the positron in 1932,
the question of the existence of antimatter in the Universe has
puzzled astrophysicists. Besides the production of positrons in
the laboratory and by cosmic rays in our atmosphere, positrons were
supposed to be produced in a multitude of astrophysical environments.

{\bf Observations} : In the seventies, balloon instruments provided first evidence for
e$^{-}$e$^{+}$ annihilation from the Galactic Center region.  As the line
was discovered at an energy of 476 $\pm$ 26 keV \cite{johnson72}, the
physical process behind the emission was initially ambiguous and had to
await the advent of high resolution spectrometers.  In 1977, germanium
semiconductors, flown for the first time on balloons, allowed to 
identify a narrow annihilation line at 511 keV, its width
being of a few keV only \cite{albernhe81}, \cite{leventhal78}.
The eighties were marked by ups and downs in the
measured 511 keV flux through a series of observations performed by 
balloon-borne germanium detectors (principally the telescopes of
Bell-Sandia and Goddard Space Flight Center).  
Those results were interpreted as the
signature of a fluctuating compact source of 
annihilation radiation at the Galactic
Center (see e.g. \cite{leventhal91}).  Additional evidence for this scenario
came initially from HEAO-3 \cite{riegler81} reporting
variability in the period between fall 1979 and spring 1980.  Yet, during
the early nineties, this interpretation was more and more questioned, since
neither eight years of SMM data \cite{share90} nor the revisited
data of the HEAO-3 Ge detectors
\cite{mahoney93} showed evidence
for variability in the 511~keV flux.  Throughout the nineties, CGRO's
Oriented Scintillation Spectrometer Experiment (OSSE) measured steady
fluxes from a Galactic bulge and disk component \cite{purc97}
and rough skymaps became available based on the combined data from OSSE,
SMM and TGRS.  The corresponding pre-INTEGRAL view of galactic positron
annihilation invokes different scenarios based on two main components - a
central bulge or halo and a Galactic disk \cite{kinz01,miln01}: 
``bulge-dominated'' models comprise a halo bulge plus a
thin disk, while in the ``disk-dominated'' scenarii a 2D Gaussian bulge without a
halo combines with a thick disk.  The integrated annihilation rate
is similar in the various models, but the data do not strongly constrain the
bulge to disk flux ratio which spans a range going from B/D = 3.3 (bulge-dominated) to
B/D=0.2 (disk-dominated).

A possible third component at positive Galactic latitude (between 
l=$9^{\circ}$ - $12^{\circ}$) was first attributed to an annihilation fountain
in the Galactic center \cite{derm97}; however the intensity and
morphology of this feature were only 
weekly constrained by the data \cite{miln01}.

Regardless of their discrepant flux estimates, pre-INTEGRAL missions were in
good agreement with respect to the observed ``positronium fraction'' f$_{ps}$, the
fraction of positrons which annihilates after having formed positronium
atoms.  The positronium (Ps) fraction is calculated as 
f$_{ps}$~=~2/[2.25($I_{511}/I_{ps}$)~+~1.5] where $I_{511}$ and $I_{ps}$ are the
intensities in the 511~keV line and the 3-photon continuum emissions
respectively.  Observed Ps fractions converged towards f$_{ps}$=0.93$\pm$0.04
\cite{harr98,kinz01}.

\begin{table*}
\label{historic measurements}
    \caption{the Galactic Center 511 keV line measured by high resolution
spectrometers}
    \begin{array}[b]{lcllll}
\noalign{\smallskip}
\hline
\hline
\noalign{\smallskip}
instrument  & year & flux & centroid & width & ref.\\
            &      & [10^{-3} &          & FWHM  &    \\
            &      & ph\ cm^{-2}\ s^{-1}]  & [keV] & [keV] \\
\hline
\noalign{\smallskip}

HEA0 3 & 1979-1980   & 1.13 \pm .13 \ (a)& 510.92 \pm 0.23 &
1.6^{+0.9}_{-1.6} & (1)\\

SMM & 1980-1986 & 2.1 \pm .4  \ (a)& & unresolved & (2)\\

GRIS & 1988,\ 1992 & 0.88 \pm .07  \ (b)& & 2.5
\pm .4 & (3)\\

HEXAGONE & 1989 & 0.95 \pm .23  \ (b)& 511.54 \pm 0.34 & 2.66 \pm
.60 &(4) \\

OSSE & 1991-2000 & 2.4 - 3.1  \ (c)&  & unresolved & (5) \\

TGRS & 1995-1997 & 1.07 \pm .05 \ (a)& 511.98 \pm 0.10 & 1.81 \pm
.54 & (6) \\

SPI & 2003 & 0.96^{+.21}_{-.14}  \ (d) & 511.02^{+0.08}_{-0.09} &
2.67^{+0.30}_{-0.33} & (7) \\

 \noalign{\smallskip}
 \hline
    \end{array}
{(a) for a point source or spatially unresolved source at the Galactic
Center; (b) Galactic Center flux within the instruments field of view :
$18^{\circ}$ and $19^{\circ}$ FWHM for GRIS and HEXAGONE, respectively; (c)
best fit fluxes of a bulge and disk model - uncertainty from thin/thick
disk model; d) best fit flux in spherical Gaussian with $8^{\circ}$ FWHM
centered the GC.  \\
(1) Mahoney {\it et al.} 1994; (2) Share {\it et al.} 1988; (3) Leventhal
{\it et al.} 1993; (4) Durouchoux {\it et al.} 1993; (5) Kinzer {\it et
al.} 2001 ; (5) Harris {\it et al.} 1998; (5) Jean {\it et al.} 2004}
\end{table*}

Since the launch of INTEGRAL in October 2002, a large part of this
mission's core program has been devoted to a Galactic Central Region Deep
Exposure (GCDE).  Imaging analyses from data of the INTEGRAL spectroenter SPI during
the first year shows the 511~keV emission to be spatially extended $(9^\circ\pm~1^\circ$ 
FWHM), however rather symmetric and centered around the Galactic Center \cite{jean511_04,weid04,chur04}.  
The corresponding bulge flux is
 10$^{-3}$~ph~cm$^{-2}$ s$^{-1}$, with a 15\% 
uncertainty being dominated by the width of the Gaussian intensity
distribution.  Marginal evidence for emission from a Galactic
disk has been found only recently \cite{knoe05}, with an intensity
which leaves little room for positrons other than the ones from \Al decay. 
However, a positive latitude enhancement of annihilation emission, as 
had been suggested from OSSE measurements, appears rather unlikely, from
INTEGRAL's measurements. The inner-Galaxy emission at 511~keV
can not be explained by a single source, but the contribution of a number
of point sources can not yet be excluded by SPI.  Spectroscopy of 511 keV
line emission from the bulge resulted in a best fit energy of
$511.02^{+0.08}_{-0.09}$ keV and an intrinsic line width of
$2.6\pm~0.3$~keV FWHM \cite{lon04,chur04}.
A positronium continuum is detected  \cite{chur04,stro04} and contains about
3--5 times the $\gamma$-ray flux than the line itself 
(a Positronium fraction of 0.93 is
consistently derived). Together with
the detailed shape of the 511~keV line, this suggests
annihilation in a warm medium \cite{chur04,gues04,gues05} (Fig.~\ref{511_lineshape}).
Annihilation in a hot medium would produce a much broader line than observed,
whereas the annihilation from a cold medium after thermalization would cause
a narrower line profile than observed. Also, a substantial contribution from 
annihilation on the surface of
interstellar grains is considered unlikely, from the line peak to wing area
comparisons. Hence, detailed 511~keV spectroscopy introduced a new probe 
of the physics of the interstellar medium in the Galactic bulge. 

\begin{figure}[ht]
\centering
\includegraphics[width=9cm]{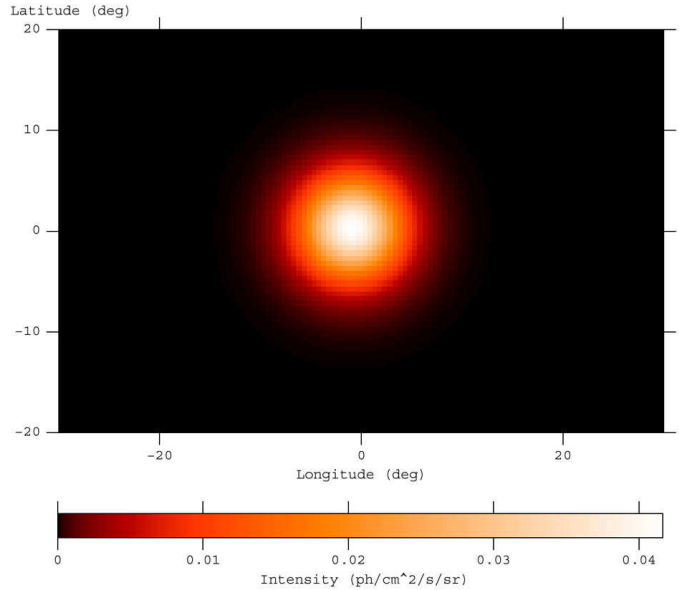}
\caption{INTEGRAL/SPI best fit model for the 511 keV line emission from the
GC region.  A Gaussian with $8^{\circ}$ FWHM, representing the Galactic bulge,
is sufficient to explain the data of the first GCDE, from \cite{weid04}.
 Recent imaging deconvolutions through various method variants essentially
 confirm such a model, yet are still too limited to allow extraction of
 significant details beyond such a model \cite{knoe05}.
}
      \label{PvB_511_map}
\end{figure}
\begin{figure}[ht]
\centering
\includegraphics[width=9cm]{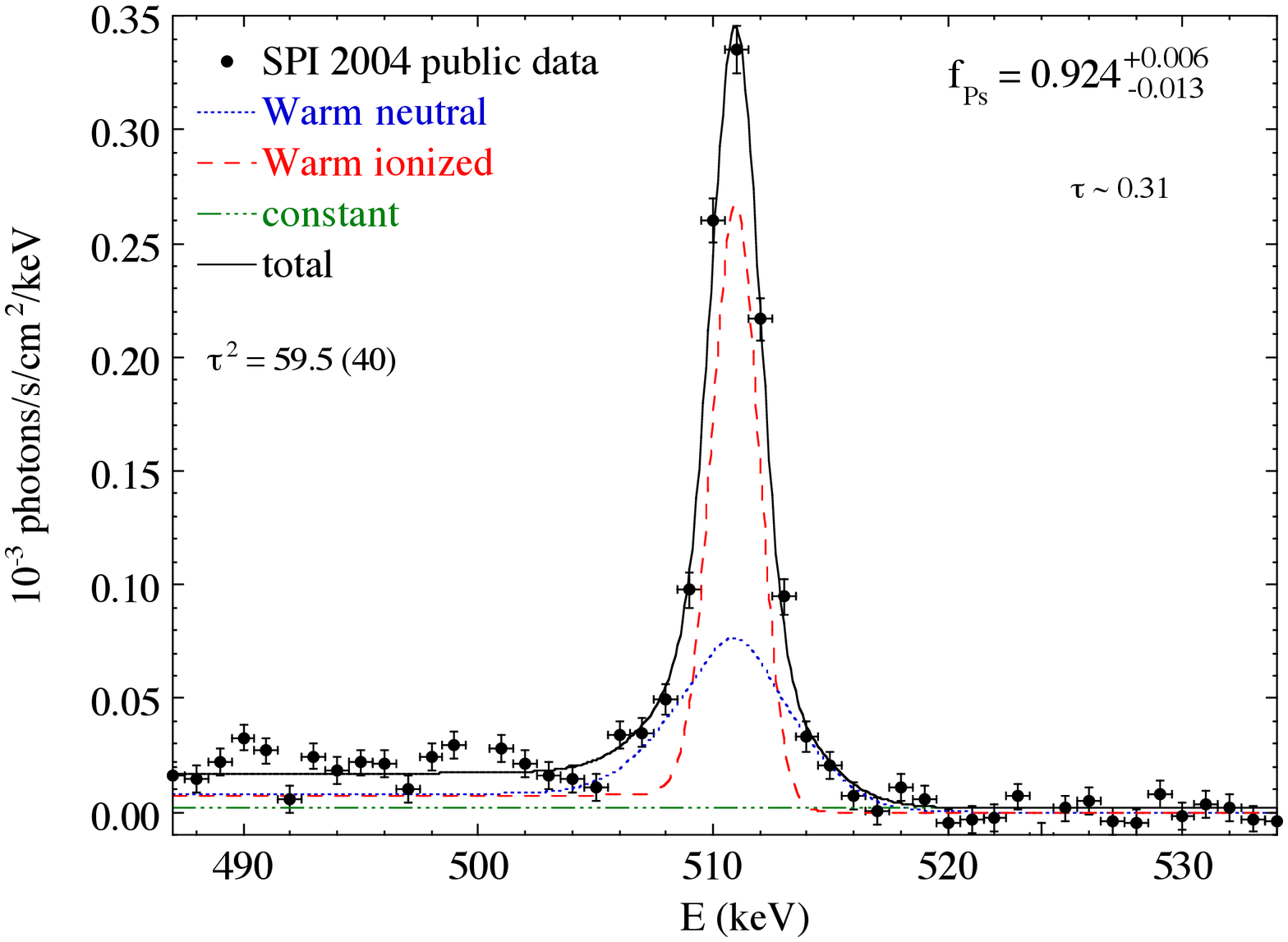}
\caption{INTEGRAL/SPI measurement of the shape of the 511~keV line
from positron annihilation. Compared to expectations from annihilations
in different phases of the ISM, one finds that warm ISM in its neutral and 
ionized parts contributes equal amounts to total annihilation, while cold 
and hot ISM phases appear insignificant \cite{chur04,jean05,gues05}.
}
      \label{511_lineshape}
\end{figure}

With its superior angular resolution and good sensitivity for point
sources, INTEGRAL's imager IBIS has the potential to reveal whether the
extended bulge emission is of genuinely diffuse origin or results from blended
emission from a number of compact sources.  The analysis of INTEGRAL/ISGRI data
during the first year of the missions 
Galactic Center Deep Exposure \cite{decesare2004} 
shows no evidence for point sources at 511 keV;
the $2 \sigma$ upper limit for resolved single point sources is
estimated to 1.6 10$^{-4}$  ph cm$^{-2}$ s$^{-1}$.

In the near future, additional exposure and improved knowledge of background
systematics will refine INTEGRAL's image of Galactic e$^{-}$e$^{+}$
annihilation, and better constrain the numerous models proposed for its origin.

{\bf Astrophysical Models} :
Our view of Galactic positron annihilation is considerably
simplified with the SPI results: 
The e$^-$e$^+$ emission apparently originates from a simple 8$^\circ$-wide (FWHM) 
central bulge, and there is not any evidence  
for deviations from such Galactocentric symmetry.
The weakness of the disk component comes as a surprise,
since positrons are necessarily produced in the galactic disk by
radioactive isotopes produced in supernovae and novae.  
The observed \Al alone (see section 3.3) can account
for $\sim$ 5 10$^{-4}$ photons cm$^{-2}$s$^{-1}$ of the disk's 511~keV
$\gamma$-ray brightness, since its
decay is accompanied by the emission of a positron in 85\% of the cases.

INTEGRAL is therefore reformulating the question on Galactic e$^-$e$^+$
annihilation : In order to explain $\simeq 10^{-3}$ ph~cm$^{-2}$s$^{-1}$ 
in the 511~keV line at distance of the Galactic Center, an
annihilation rate of $\simeq 10^{43}$ positrons per second is required in a
relatively small bulge region around the GC (assuming a distance of 8~kpc to the GC and a
positronium fraction of f$_{ps}$=0.94, resulting in 0.59 511~keV line photons 
emitted per positron annihilation).

Amongst the mechanisms that have been proposed for the origin of Galactic positrons
are 
(a) $\beta^+$-decaying radioactive isotopes, 
(b) the decay of
$\pi^{+}$'s produced in cosmic-ray proton interaction with interstellar
nuclei, 
(c) high energy processes (e.g. pair creation by high-energy photons) 
in compact objects, and 
(d) bosonic dark matter
annihilation of  low mass particles (1-100 MeV) in the Galactic halo;
the latter has been specificaly proposed to explain the 
morphology of the 511~keV emission after the first SPI measurmenets.

The simple  morphology of the 511~keV emission observed by SPI, if conservatively
interpreted, suggests that the Galaxy's old stellar population is
at its origin; alternatively, something much more exotic, like dark matter,
may be at work. In the following we  focus on e$^+$ sources concentrated in 
the central region, and neglect \Al and cosmic ray production of positrons
(which should be clearly seen in the disk).

{\it Type Ia supernovae} : The light curves of SNe Ia are powered by the deposition
in the expanding SN ejecta of the gamma-rays and positrons produced by
various radioactivities, especially the decay of $^{56}$Co, $^{44}$Ti
and $^{57}$Co ( the latter does not release positrons). 
Whether the positrons escape from the ejecta or not, 
depends on the strength and geometry of the
magnetic field.  Positron transport in SNIa models was simulated \cite{chan93}.
It was found that, for favorable magnetic field
configurations, $\sim$ 5\% of $^{56}$Co decay positrons may escape the ejecta.
Comparing a sample of
late light curves of supernovae Type Ia with simulations \cite{milne_SN99,milne_SN01}, 
the underlying idea can be tested that
positron escape leads to power loss from the system and a drop in
the late lightcurve.  These studies conclude that the number of positrons
escaping a typical SNIa is $N \sim 8^{+7}_{-4}$ 10$^{52}$, or about
3\% of the total amount of positrons released by \Co \ decay.
Note that the amount of \Ti produced in SNIa is 4-5 orders of magnitude
smaller than the one of \Ni \cite{iwam99}, 
so that even if all the positrons released
by  \Ti decay escape the supernova, their contribution to the above
number N is negligible.

The production rate of positrons R from SNIa in the
bulge might then be estimated as R = M F N \cite{prantzos_iws04}, 
where M is the mass of the
bulge (in units of $10^{10}$\ms) and F is the SNIa frequency in a
bulge-like system (expressed as number of SNe 
per century and per 10$^{10}$ \ms \ in stars).  
Adopting M=$1.5\pm$0.5 \cite{launhardt02} and 
F=$0.044\pm 0.03$ \cite{Mannucci04} 
a positron production rate R $\simeq 2 \ 10^{42}$ positrons per second is
obtained. This figure corresponds to about 20 \% of the required
bulge positron production rate after SPI and is twice as large as
in previous estimates (e.g. \cite{prantzos_iws04}), because of the recently revised
SN rates of \cite{Mannucci04}. This increase with respect to previous estimates, 
although not solving the problem, should remind us that the uncertainties 
(statistical and systematic) of the various
parameters entering the calculation of R may be substantially
larger than formally quoted. It appears to us that current
uncertainties are such that SNIa can either be major contributors to the
511 keV emission or have a negligible contribution.

On the one hand, the SNIa frequency F 
may still be considerably underestimated, 
due to dust extinction in the inner galaxian regions and to the difficulty of
detecting supernovae against the bright background of stellar bulges;
indeed, it is intriguing that no SNIa have been found up to now in the
bulges of spirals (\cite{wang97}).
On the other hand, the estimate of the number of positrons N by 
Milne et al. \cite{milne_SN01}  
might be too optimistic. For one well studied case, the SN Ia 2000cx, 
Sollerman {\it et al.} \cite{sollerman04} 
find that the decay of the late optical lightcurve is accompanied by an
increased importance of the near-IR emissivity 
and does not necessarily imply 
an escape of  positrons from the ejecta. In the simple
model of Sollerman {\it et al.} accounting for the UVOIR observations,
all positrons are trapped in the ejecta. The investigation of positron escape from
SNIa through the study of late lightcurves requires more
observations and  more realistic models. The contribution of this class of objects
appears promising, but other sources should definitely be sought.

{\it LMXB} : The morphology of their distribution makes Low Mass X-Ray
Binaries (LMXBs) very promising candidate sources \cite{prantzos_iws04}: their observed distribution
in the Milky Way is strongly concentrated towards the Galactic bulge.
Furthermore, the collective X-ray emissivity of Galactic LMXBs is 10$^{39}$
erg/s \cite{grimm02}, compared to
the  10$^{37}$ erg/s required to produce
$10^{43}$ positrons per second.
It thus is sufficient to convert only 1\% of the available energy into
positrons to explain the bulge emission.  However, the mechanism of that
conversion is not known yet.

{\it Light dark matter} : It has been argued (\cite{boeh04}) that low mass bosonic dark
matter may be the source of the observed 511 keV emission line.  The 
dark matter particles annihilate throughout the Galactic
bulge into e$^{-}$e$^{+}$ pairs which, after deccelerating,
annihilate into 511 keV photons. The proposed particles are quite
light (in the 1-100 MeV range) so that their annihilation does not
produce undesirable high energy gamma-rays, and in that respect
they do not correspond to the most commonly discussed dark
matter candidate, which has mass in the GeV to TeV range. Moreover, rather
special properties are required for such light particles to
justify why they have  escaped detection up to now in accelerators
such as the LEP. 

Various profiles for the dark matter density have been proposed,
differing vastly from each other.  It is likely that the dark matter
density profile is cusped as 1/r at small galactic radii; hence the gamma ray flux
would be expected to be considerably enhanced.  Discriminating 
between ill-defined dark
matter distributions and other bulge candidate sources 
(LMXB, SNIa)  with INTEGRAL/SPI will be a very difficult task.
More generally, it is hard to evaluate the plausibility
of the dark matter  hypothesis, since the required properties of the source
(i.e. density profile, annihilation cross-section) are completely
unknown/unconstrained; in fact, the observed properties of the
511~keV emission (intensity and density profile) are used in
\cite{boeh04} in order to derive the properties of the
dark matter source of positrons.

Another method to determine whether the 511 keV line is due to dark matter
annihilation is to seek a 511 keV signature from low surface
brightness dwarf galaxies \cite{hoop03}, which appear to be dark
matter-dominated.  If the emission line detected in our Galaxy
is due to dark matter annihilation, then a relatively intense 511 keV line from
nearby dwarf galaxies is also expected.  For the Sagittarius dwarf galaxy, a 511
keV flux of (1-7) 10$^{-4}$ photons cm$^{-2}$s$^{-1}$ was prediced
\cite{hoop03}.  Based on the limited statistics of INTEGRAL/SPI's first year core
program data, a 2$\sigma$ upper
limit of $2.5 \times 10^{-4}$ photons cm$^{-2}$s$^{-1}$ on the annihilation
flux from the Galaxy is set \cite{cordier04}.

However, at this point, the non-detection of the Sagittarius dwarf galaxy can neither rule out nor
confirm the light dark matter hypothesis. Only with deeper exposure of
dark matter targets (such as Sagittarius or the globular cluster
Palomar-13), SPI/INTEGRAL will reach adequate sensitivity to
efffectively reject or confirm light dark matter annihilation as source
of the bulge 511 keV emission.


\subsection{Nuclear De-excitation Lines: \\
   He Isotopic Abundances in the Sun, and Particle Acceleration Physics}

Cosmic rays and high-energy $\gamma$-rays from active galactic nuclei (AGN),
supernova remnants, and the Galactic interstellar medium  demonstrate
the existence of efficient particle accelerators in the universe \cite{stro99,torr03}.
Although it is generally believed that Fermi acceleration is the only
mechanism capable to provide the observed particle energies,
the acceleration process, and the associated injection of suprathermal
particles into the acceleration region, is far from being understood.
Energetic particles produce characteristic $\gamma$-rays from nuclear
de-excitation upon their collisions with ambient matter.

The Sun is our closest laboratory for the study of energetic particles.
Within the Solar System, solar energetic particles can even be directly measured
through particle detectors in interplanetary space\cite{cohe99}, although
their interpretation is complex due to modulation from local magnetic fields.
Characteristic $\gamma$-ray emission is produced when solar
flare events produce a burst of energetic particles, which collides with
gas of the upper solar atmosphere.

Detailed measurements of $\gamma$-ray flare spectra have been obtained by
the Compton Observatory at rather modest resolution \cite{murp97}, and more recently
with Ge-detector resolution by the RHESSI experiment \cite{lin03}.
Such spectra exhibit complex superpositions of narrow solar-atmosphere and
broad solar-flare particle de-excitation lines, often burried in an
intense electron Bremsstrahlung continuum \cite{murp97}.
There is correlated and slower variability of lines originating from neutron
interactions, as compared to other nuclear deexcitation lines.
This difference between lines caused by high-energy proton spallations as compared
to lines caused by low-energy
proton collisions and to bremsstrahlung from flare electrons  suggests different 
acceleration sites for the high- and low-energy particles which hit solar atmosphere
material in such flares:
Fermi acceleration initiated by large-scale processes in the solar magnetosphere may
lead to very energetic solar-flare particles,
far beyond electrostatic or Fermi accelerators set up by more rapid
loop reconnection events in the loop structures, which provide the electron and low-energy
particle components \cite{huds95}.

With RHESSI's high spectral resolution, specific $\gamma$-ray lines could
be studied in much more detail, investigating nuclear-physics processes.

The positron annihilation line at 511~keV provides a unique opportunity to
investigate how positron annihilation proceeds: Positrons are produced at
high energies, both from nuclear interactions and from radio-isotopes produced
through spallation reactions. Thermalization mostly preceeds annihilation,
which then can occur on bound electrons to produce a rather narrow line
(FWHM \about~1.5~keV), while annihilation through formation of positronium
atoms not only produces the triplet-state continuum spectrum below 511~keV,
but also a rather broad 511~keV line (FWHM \about~6-7~keV) \cite{buss79,shar03a}.
If annihilation occurs before thermalization, or else in a hot environment, also
the charge exchange reaction channel may produce a broader line. In this respect,
RHESSI's findings are puzzling \cite{shar03}: The measured broad line
would suggest annihilation through Positronium formation, but the
line-to-annihilation-continuum ratio is inconsistent with this explanation;
on the other hand, a thermal interpretation of the broad line violates the
sequence of thermalization and charge-exchange annihilation, because it suggests
annihilation high up in the solar atmosphere, well above the positron production
region.

In impulsive solar flares, isotopic enrichment of $^3$He up to \about~4 orders
of magnitude above the ratio in solar wind of 5~10$^{-4}$
has been observed; this is much discussed, especially in terms of isotopically-selective
particle acceleration. The reaction $^{16}$O($^3$He,p)${18}$F$^*$ produces
characteristic $\gamma$-ray lines at 937, 1042, and 1081~keV, which allows
$^3$He abundance determination through solar flare $\gamma$-ray spectra
\cite{mand97}. From SMM and CGRO flare $\gamma$-ray spectra, flare-averaged $^3$He/$^4$He
abundance ratios between 0.1 and 1 have been deduced in this way \cite{shar98}. However,
uncertainties from the above reaction rate and its energy dependence are a concern \cite{tati03}.

The geometry of the accelerated-particle beam is reflected in a variety
of $\gamma$-ray line parameters:
RHESSI's high imaging resolution through a rotating modulation collimator
allowed imaging in the 2.223~MeV neutron capture line, finding it offset
from the source of X-ray bremsstrahlung continuum \cite{hurf03}; this suggests
that the flare particles hit the solar atmosphere at an angle and not
perpendicularly.
RHESSI and SPI found significant redshifts in $\gamma$-ray lines from C, O,
Ne, Mg, Si, and Fe \cite{smit03f,gros04}, which suggest downward-beamed
nuclei being responsible for the emission.
$^7$Be and $^7$Li lines at 429 and 478~keV, respectively, from collisions
of flare and ambient $\alpha$ nuclei are, however, so broad that isotropic
rather than beamed energetic $\alpha$-particles must be assumed \cite{shar03}.

Evidently, more RHESSI flare $\gamma$-ray observations will be necessary to
obtain a more consistent picture on particle acceleration in solar flares.

The complexity of nuclear de-excitation line
$\gamma$-ray spectra is expected to be greatly simplified in the "thin-target"
configuration, which presumably is realized when low-energy
cosmic rays collide with ambient matter in star-forming regions and the general
interstellar medium \cite{rama79}. The COMPTEL discovery of intense
nuclear de-excitation $\gamma$-rays \cite{bloe94} came as a surprise and provided a
great stimulus to studies of low-energy cosmic rays; the experimenters
withdrew their discovery however, when they noted that instrumental
background may have caused such a signal artifact \cite{bloe99}.
Thus, nuclear lines from cosmic ray interactions still
are to be discovered. Predicted intensities \cite{higd87} of
$\simeq$10$^{-6} $\flux\ leave little hope for INTEGRAL.


\section{SUMMARY AND PERSPECTIVES}

Despite the serious handicaps of a rather modest spatial resolution and
signal/noise ratios in the percent regime, astronomy with gamma-ray lines
  became a major discipline
of modern astrophysics in the 90ies.
In particular, this branch of astrophysics provides unique views to:
\begin{itemize}
\item{}
the interiors of supernovae, through measurements of $^{56}$Co 
and $^{57}$Co as detected in SN1987A, and of $^{44}$Ti as detected in Cas~A;
\item{}
the curent large-scale star formation and massive-star/ISM interactions 
in the Milky Way, through  mapping and spectroscopy of $^{26}$Al in the Galaxy;
\item{}
stellar nucleosynthesis, through the constraints imposed by the 
 abundance ratios of $^{44}$Ti/$^{56}$Co and $^{26}$All/$^{60}$Fe;
\item{}
the source (and, perhaps, propagation) of positrons in the
 Galaxy, through the intensity and morphology of the 511 keV line, as mapped
 by SPI/INTEGRAL
\end{itemize}
 Those important results have lead to several new questions: 
\begin{itemize}
\item{}
What fraction of radioactive energy is converted into other forms
of energy in supernovae? (The issues are: What is the absolute amount
of (presently indirectly-inferred) radioactive \Ni in SNIa, and of
\Ti in core-collapse SNe; what is the magnitude of positron leakage from
supernovae; is the morphology of expanding
supernova envelopes dominated by inhomogeneities, ``bullets'', 
filaments, jets, magnetic fields)?
\item{}
How good are our (basically one-dimensional) models for nova and supernova
nucleosynthesis, in view of important 3D effects such as rotation and
convective mixing? (The issues are: How much \Ti mass can in principle be 
(and is effectively) ejected from regions near the mass cut between compact remnant and
ejected supernova envelope; how realistic are nova \Na yields, what are
the ejected nova-envelope masses, how critical are the seed compositions
for explosive hydrogen burning in novae.)
\item{}
What is the dynamic range of physical conditions expected for nucleosynthesis
events? (The issues are: How does nucleosynthesis vary with
 stellar mass and metallicity,
 what are the applicable supernova rates in specific ensembles of massive stars; 
 how much clustering of events
 is there in space and time; does self-enrichment play a major role;
 is star formation triggered in dense, active nucleosynthesis regions)
\item{}
What is the interplay between interstellar medium and energy outputs and material
ejecta from massive stars?
(The issues are: Which are the characteristic
temporal and spatial scales of energy and material flows in
interstellar gas; how does the morphology of the interstellar
 medium affect such scales; do lower and higher mass stars form at special
 epochs and locations; where are the presolar grains formed, and how 
 are they processed before we detect them in the meteoritic laboratory?)
\item{}
How do positrons end up annihilating to produce $\gamma$-ray photons?
(The issues are: How do they escape their sources, how do they propagate
through interstellar space and possibly the Galactic halo; what are the
resulting slowing down times, and their total lifetimes before annihilation,
and what is its spread; what is the final annihilation environment, and
how localized does annihilation occur; what is the ratio between the different
sources of nucleosynthesis, compact stars, and possibly dark matter?
\item{}
How are particles accelerated to cosmic-ray energies?
(The issues are: Which isotopes are selectively injected into the
 acceleration region; how does ``injection'' occur, from thermalized seed
 particles; which magnetic-field configuration sets up accelerators at
 different sites and on different scales; how do these accelerators
 shape the isotopic composition; which secondary isotopes are produced
 by cosmic ray spallation reactions; where and how is spallation
 nucleosynthesis most efficient; what is the role of compact stars
 and their accretion?

\end{itemize}

With the INTEGRAL \cite{wink03i,vedr03,roqu03} and RHESSI \cite{lin02}
space experiments, now the new step is taken with high-resolution
spectrometers, improving sensitivities by \about~an order of magnitude
through the use of large volumes of Ge detectors.
More importantly, with these instruments we now can observe
kinematic signatures from Doppler-shifted energy values in
expanding/accelerated radioactive material and in positron annihilations
in interstellar space through different reaction channels.
With such measurements, new valuable constraints are being obtained on sources
of cosmic nucleosynthesis.

\end{document}